\documentclass[11pt,a4paper]{article}
\usepackage{graphicx} 
\usepackage{amsmath} 
\usepackage{amsfonts} 
\usepackage{subfig}
\usepackage{latexsym}

\begin{document}
\title{Entanglement between Demand and Supply\\ in Markets with Bandwagon Goods}

\author{Mirta B. Gordon {\small (1)}, Jean-Pierre Nadal {\small (2,3)}, \\Denis Phan {\small (4)}, Viktoriya Semeshenko {\small (1,5)}}
\vspace{2cm}
\date{{ \small
(1) Laboratoire d'Informatique de Grenoble (UMR 5217 CNRS -- INRIA -- Grenoble INP -- UPMF -- UJF), Grenoble, France \\
(2) Centre d'Analyse et de Math\'ematique Sociales (UMR 8557 CNRS -- EHESS), 
Ecole des Hautes Etudes en Sciences Sociales, Paris, France \\
(3) Laboratoire de Physique Statistique (UMR 8550 CNRS -- ENS -- UPMC -- Universit\'e Paris Diderot), 
Ecole Normale Sup\'erieure, Paris, France\\
(4) Groupe d'Etude des M\'ethodes de l'Analyse Sociologique de la Sorbonne (UMR 8598 CNRS -- Paris IV), Universit\'e Paris Sorbonne-Paris IV, Paris, France \\
(5) Facultad de Ciencias Econ\'omicas, 
Universidad de Buenos Aires, Buenos Aires, Argentina \\
$\;$}\\
{\small Version 28 Nov. 2012, accepted for publication in\\Journal of Statistical Physics. The final publication is available at link.springer.com, doi:10.1007/s10955-012-0660-1}}

\maketitle

\begin{abstract}
Whenever customers' choices (e.g. to buy or not a given good) depend on others choices (cases coined 'positive externalities' or 'bandwagon effect' in the economic literature), the demand may be multiply valued: for a same posted price, there is either a small number of buyers, or a large one -- in which case one says that the customers {\it coordinate}. This leads to a dilemma for the seller: should he sell at a high price, targeting a small number of buyers, or at low price targeting a large number of buyers? In this paper  we show that the interaction between demand and supply is even more complex than expected, leading to what we call the {\it curse of coordination}: the pricing strategy for the seller which aimed at maximizing his profit corresponds to posting  a price which, not only assumes that the customers will coordinate, but also 
lies very near the critical price value at which such high demand no more exists. This is obtained by the detailed mathematical analysis of a particular model formally related to the Random Field Ising Model and to a model 
introduced in social sciences by T C Schelling in the 70's.\\
$\;$
\end{abstract}

\pagebreak
\section{Introduction}
\label{sec:introduction}

Interactions play a crucial role on the collective outcomes observed in social systems. The decision of leaving a neighborhood \cite{Schelling71}, to attend a seminar \cite{Schelling73,Schelling} or a crowded bar \cite{Arthur94,Arthur99}, to participate to collective actions such as strikes and riots \cite{Granovetter78}, are particular examples where the emergent state may be at odds with widespread individual wishes. It has been further suggested that social interactions may explain the school dropout \cite{Crane91}, the persistence in the educational level within some neighborhoods \cite{Durlauf96}, the related consequences in the stratification of investment in human capital and economic segregation \cite{Ben96}, the large dispersion in urban crime through cities with similar characteristics \cite{GlaeserSacerdoteScheinkman96}, the emergence of social norms \cite{Ostrom00}, the labor market behavior and related unemployment patterns \cite{Topa01,ConleyTopa02}, the housing demand \cite{IoannidesZabel03}, 
the existence of poverty traps \cite{Durlauf06}, the smoking behavior \cite{Krauth06a,Krauth06b,SoeteventKooreman06}, etc. In all these cases, social interactions give raise to multiple equilibria.

In economics there is a growing literature that recognizes the influence of social interactions in market situations like the subscription to a telephone network \cite{Art73,Roh74,vRab74,Cur87} or the choice of a computer operating system \cite{KatzShapiro94}, etc. The maximum price that each consumer accepts to pay for a given good (the {\it willingness to pay}), depends not only on his/her idiosyncratic preferences but also on the choice made by others \cite{ShapVar99,Rohlfs01}. If the social interaction (called {\em externality} in economics) is positive, the 
willingness to pay for the most popular good increases. 
This leads to a so called {\it bandwagon effect} where this good is bought even by individuals who otherwise would have never made this choice. 
General aspects of bandwagon effects on the consumers demand 
have been discussed in the economics literature \cite{Leibenstein50,GranovetterSoong86,BeckerMurphy00,Manski00,Rohlfs01,Durlauf97,BrockDurlauf01a,NaPhGoVa05}. The case when the willingness to pay for a good increases proportionally to the number of buyers, has received much attention. One shows that the demand curve (the relationship between the price and the quantity of goods that the customers are willing to buy) may be non monotonic. There is then a range of prices for which there are multiple market equilibria. The highest demand is Pareto-optimal, that is, at a given price, it satisfies the largest number of customers. If this high demand equilibrium is achieved, in economics one says that the customers {\em coordinate} (a large number of customers make the same choice which is at their common benefit). 
Like a physical system trapped in a metastable state, the Pareto-optimal equilibrium may not be achieved. In economics, it is then said that there is failure of coordination among the customers. 

The particular model we consider here, and to be described more precisely in the next section, corresponds to the case of binary choices (to buy or not to buy a single unit of an homogeneous not-divisible good), with the social influence contributing additively to the willingness to pay, and heterogeneous agents in their idiosyncratic willingness to pay. 
Although we consider a market context, this model has a wide range of applications in social (non economics) situations. It 
is a variant of the ``Dying seminar'', one of the models introduced by the social scientist T. C. Schelling in the 70's\cite{Schelling73,Schelling}, in order to explain how individual actions may combine to produce unanticipated social collective outcomes. Soon after Schelling, the same model has been notably considered by M. Granovetter \cite{Granovetter78} for the formation of riots. This model is formally related to the Random Field Ising Model (RFIM) \cite{Sethna} at zero temperature, as shown in \cite{NaPhGoVa05} -- the Ising spin corresponding to the binary choice of an agent, the  exchange constant to the social coupling, and the quenched random field to the idiosyncratic part of the willingness to pay. Actually, there is already an important literature making use of the RFIM and other Ising type models for addressing issues in social and economic science. The first explicit use of an Ising model in economics can be traced back to the work of the mathematician Follmer in the mid 70's \cite{Follmer74}, 
and the first use of the RFIM in social science to the 1982 paper of Galam {\it et al} \cite{GaGeSh}. Since then such statistical physics models have been exploited by both economists and physicists, see e. g. \cite{Orlean95,Durlauf97,WeiSta} and the review papers \cite{PhaGorNad04,BouchaudReview2012}. 
 
We have extensively studied this Schelling/RFIM model \cite{NaPhGoVa05,GoNaPhSeM3as} in the mean field case, as a function of the characteristics of the willingness to pay distribution. As expected, the existence, the number and the nature of multiple equilibria are generic properties:
 they only depend on the strength of the social coupling and on the number of maxima of the willingness to pay distribution, but not on its details.  We have shown that some specific properties of the phase boundaries in parameter space are even universal: they are quantitatively, and not only qualitatively, identical for a wide family of different random field distributions. Interestingly, empirical analyses of some collective social behaviours have been shown to
exhibit scaling laws predicted by the critical properties of this model \cite{MichardBouchaud,BorghesiBouchaud}. 
The model has also been extended to a dynamical setting with adaptive agents who update their believes, or equivalently their willingness to pay, according to the current market state \cite{WeiSta,SeGoNa08}.

Up to now we only mentioned the customers behaviors 
facing a posted price. The analysis of the possible strategies of a seller (the offer or {\em supply}) with customers under social influence 
has deserved much less attention. 
Within the economics literature, Granovetter and Soong \cite{GranovetterSoong86} have discussed  
possible pricing strategies, i.e. what price to post in order to maximize the seller's profit,
within the same context of positive externalities. However, their analysis  remains qualitative and notably does not explore in details the consequences of 
multiple equilibria on pricing. 
A particular insightful paper is Becker's note \cite{Becker91} attributing to social interactions the fact that popular restaurants do not increase their prices despite a persistent excess demand.

In this paper, we go one step further in the analysis by considering a seller without competitors (monopoly market) facing the customer population described with the Schelling/RFIM model evoked above.  Quite interestingly, the questions of interest in social and economic contexts do not all have an interpretation in the physics context of an Ising type model. This is the case for the issues addressed in the present paper whose focus is on the pricing strategy of a seller. The single seller case corresponds to markets such as high tech ones (e g the IPhone, Nespresso machine...), where patents or other entry costs lead to a monopoly situation. 

We analyze in its generality the seller's profit optimization program.
Thanks to the linear structure of the surplus (the difference between the willingness to pay and the actual price), we are able to characterize analytically the equilibria for 
any unimodal distribution of the customers' willingness to pay. 

Having an explicit mathematical model of both demand and supply allows us to uncover market characteristics that are specific consequences of the social interactions. The relevant parameters are the strength of the social interactions, the customers mean willingness to pay, and the strength of the `disorder' (the standard deviation of the willingness to pay distribution).
A convenient normalization allows us to discuss the results on phase diagrams in a two-parameters space. We determine the bounds of the regions corresponding to qualitatively different equilibria  
and we discuss the consequences of possible price settings in each region. 

Obviously, the existence of multiple equilibria for the demand leads to the coexistence of multiple strategies for the seller. Within this domain, there is a first order transition: the optimal pricing strategy shifts from posting a high price 
attracting a small fraction of the population, to 
posting a much lower price  
to capture a larger share of the market. More surprisingly, this coexistence domain and the corresponding transition line extend into the region where the demand is single valued.

As anticipated by Becker, when the optimal strategy is associated with a high demand (requiring customers coordination), 
there is the risk that a low demand equilibrium 
exists for the same price. Our results
show that this is the generic case, that is, 
there is a very large range of parameters for which  
posting the optimal price bears a risk, because getting the optimal profit requires that customers coordinate.

Another striking result is that the optimal price is generically 
just slightly below the 
price at which the high demand equilibrium disappears. A small change in the customers characteristics may lead to a decrease of this critical price. If this change is not anticipated by the seller, the posted price may become larger than the critical price, resulting in a collapse of the demand.

The paper is organized as follows. In section \ref{sec:customers} we describe the generic properties of the model with positive-externalities and heterogeneous customers, in the case where the idiosyncratic willingness to pay has a smooth unimodal distribution. 
We assume that the monopolist is aware of the customer system properties, and is left with the problem of fixing the price. The profit optimization is discussed  in section \ref{sec:monopolist}. In section \ref{sec:monopolist_phase_diag}  we summarize our results on a phase diagram, where regions corresponding to different economic situations are represented on a plane whose axis are the 
parameters of the customer system. The possible strategies for the monopolist are discussed in section \ref{sec:discussion}. In section \ref{sec:conclusion} we summarize the results. 

\section{Customers under social influence}
\label{sec:customers}

\subsection{The Demand}
\label{sec:demand_function}
We consider a large number $N$ of  
customers that must decide whether to buy or not a single unit of a homogeneous good at the price $P$ posted by the (single) seller.
We assume that the customers' {\em idiosyncratic willingness to pay} (IWP)  
is distributed among the population according to a probability density function (pdf) of  mean $H$ and standard deviation $\sigma$, with $0<\sigma<\infty$. Upon buying at price $P$, 
agent $i$ ($i=1,...,N$) with IWP $H_i$ gets a 
surplus ${\cal S}_i$ 
linear in the fraction $\eta$ of buyers in the population: 
\begin{equation}
\label{eq:u_i}
{\cal S}_i =  H_i + J \eta - P,
\end{equation}
where $J>0$ is the strength of the social interactions. For agent $i$ the quantity ${\cal S}_i$ is the {\it ex-ante} expected surplus: what would be the outcome if he buys. We assume rational buyers that maximize their (actual) surplus: agent $i$ wants to buy if ${\cal S}_i \ge 0$, but not when ${\cal S}_i<0$ (the {\it ex-post} surplus being then ${\cal S}_i$ in the first case, and $0$ in the second case).\\
Quantities of interest are defined up to an arbitrary scale. It is convenient to make use of the following normalized parameters\footnote{In cases not studied here, $\sigma=0$ (homogeneous IWP distribution), or $\sigma=\infty$ (fat tails), one would measure quantities in units of either $J$ or $H$.}, hereafter denoted by low-case letters:
\begin{equation}
\label{eq:reduced_var}
j \equiv \frac{J}{\sigma}, \;\;\;\;
h \equiv \frac{H}{\sigma}, \;\;\;\;
p \equiv \frac{P}{\sigma}, \;\;\;\;
\hat p \equiv \frac{P - H}{\sigma} \; = p - h,
\end{equation}
With these definitions, $i$'s normalized IWP is $h_i \equiv H_i/\sigma = h+x_i$, where $x_i$ is distributed according to a pdf $f(x)$ of zero mean and unitary variance. 
We are interested in the large $N$ limit, in which case the fraction of buyers is given by the 
fixed point equation:
\begin{equation}
\eta =\int_{-s}^\infty f(x) dx = 1 - F(- s),
\label{eq:fp}
\end{equation}
where $F$ is the cumulative IWP probability distribution, and 
\begin{equation}
\label{eq:s} 
s(j,\hat p;\eta) \equiv j \eta - \hat p .
\end{equation}
Note that $s$ depends\footnote{In the argument of a function, we make use of "$;$" to separate quantities which have to be considered as parameters, here $j$ and $\hat p$, from those which appear as variables, here $\eta$.} on $p$ and $h$ {\em only} through their difference $\hat p$. This remark is 
relevant for the next section, when dealing with the pricing problem. In the game theoretic terminology, the (stable) solutions of (\ref{eq:fp}) are {\it Nash equilibria}: each agent is playing 'against' the population; within a Nash equilibrium, given the (collective) value $\eta$, no agent would have a larger payoff (surplus) by changing his choice.

As mentioned in the introduction, the above model is formally equivalent to the 'Dying seminar' model of T. C. Schelling \cite{Schelling73,Schelling}.  In this model, every agent $i$ decides to join  or not (a regular seminar, a club, a riot...) if the number of participants $N\eta$ is larger than some idiosyncratic threshold, which is the analogous, in our model, to the combination  
$N(P-H_i)/J$. This model is also equivalent to a Random Field Ising Model (RFIM) \cite{Sethna} at zero temperature, with $H_i-P$ playing the role of the quenched random field and $J$ the one of the coupling strength.  

For small $j$ 
or high $h$, 
equation (\ref{eq:fp}) has a unique solution: 
the inverse demand function\footnote{In economics, the demand curve denotes the graph {\it price  vs. quantity}, but it is meant to represent the demand function, the quantity as function of the price -- hence  the {\it inverse demand function} denotes the price as function of the quantity.}, that is the function $\hat p(\eta)$, exhibits a classical decreasing behavior 
(the smaller the price the larger the demand), like for systems without externalities.  
However, for low enough $h$, 
if $j$ is larger than some critical value $j_B$,  
equation (\ref{eq:fp}) may have multiple solutions. 
Thus, when social interactions are strong enough, $\hat p(\eta)$
has the characteristic ``down-up-down" shape assumed by Becker \cite{Becker91} and others after him. 
These two regimes are illustrated on Fig. \ref{fig:inversedemand_D}. 

In all this paper, the analysis is done when the willingness to pay has a smooth unimodal pdf $f(x)$  
whose support is the full real axis ($f'(x)=0$ only at the maximum and in the asymptotic limits $x\rightarrow \pm \infty$). 
All the numerical illustrations correspond to a logistic distribution of cumulative function $F(x)=(1+e^{-\beta x})^{-1}$, where we set $\beta=\pi/\sqrt{3}$ in order to have a unitary variance. Notice however that for the general results presented hereafter we need neither assume $f$ to be symmetric nor that its maximum is at $x=0$.

\subsection{Main steps for the analysis of the demand}
\label{sec:demand}
For completeness and later use in the analysis of the pricing problem, the main ingredients for the analysis of the demand are presented here.  
The full detailed analysis, including generalizations to other types of pdf, in particular to multi-modal ones and distributions with fat tails, may be found in \cite{GoNaPhSeM3as}. 

As briefly exposed below, the analysis leads to the introduction of two key functions of $\eta$, to be denoted $\Gamma$ and ${\cal D}$,  and we will see in the next section that analogous functions, to be denoted $\widetilde \Gamma$ and $\widetilde{\cal D}$, appear in the analysis of the seller's problem, playing the same central role in the analysis of the supply.

Since equation (\ref{eq:fp}) may have several solutions, it is convenient to consider the inverted equation, introducing the function $\Gamma$:
\begin{equation}
\label{eq:Gamma} 
\eta = 1-F(-s)  \; \Longleftrightarrow \; s = \Gamma(\eta).
\end{equation}
The properties of the IWP distribution which matters are those of the function $\Gamma$. This function $\Gamma$ is always a single valued function: it increases monotonically\footnote{Here from $- \infty$ to $+\infty$. In the case where the IWP is defined on an interval so that the normalized variable $x$ 
lies in $[x_m,x_M]$, $\Gamma$ takes the finite values $\Gamma(\eta=0) = -x_M$ and $\Gamma(\eta=1) = -x_m$.}  when $\eta$ goes from $0$ to $1$. For a unimodal distribution, $\Gamma$ has a single inflexion point. 

Replacing $s$ in the r.h.s. of (\ref{eq:Gamma}) by its expression (\ref{eq:s}) yields
\begin{equation}
\label{eq:p-h} 
\hat p={\cal D}(j;\eta)
\end{equation}
where 
\begin{equation}
{\cal D}(j;\eta) \equiv j \eta - \Gamma(\eta), 
\label{eq:D} 
\end{equation}
depends on $j$ but not on $\hat p$. 
The demand equilibria $\eta^d(\hat p,j)$ are the solutions to (\ref{eq:p-h}). 

\begin{figure}
\centering
\includegraphics[width=0.60\textwidth]{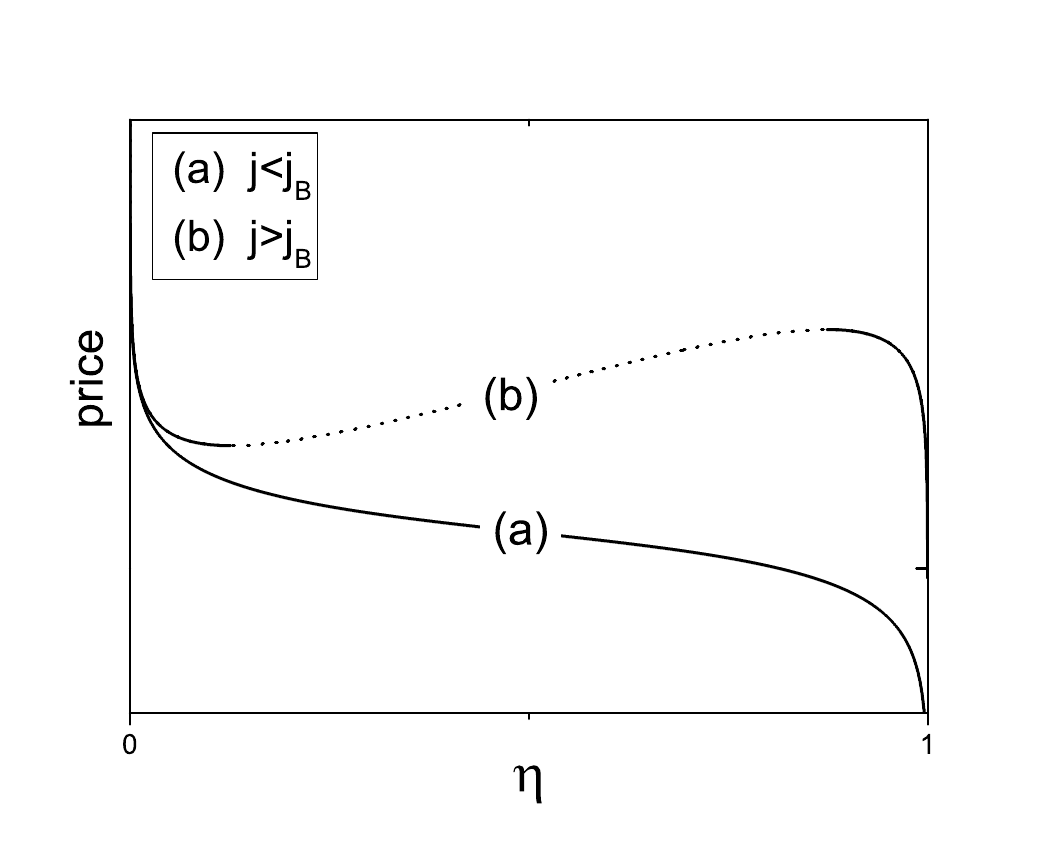}
\caption[]{\em Demand curve (price vs. fraction of buyers) for different values of the (normalized) social strength $j$ in the case of a logistic distribution of the idiosyncratic willingness to pay (IWP):
 (a) classical behaviour at low social influence or high mean willingness to pay; (b) multiply valued function at large social influence $j>j_B$ and low enough mean willingness to pay (dashed: unstable equilibria).}
\label{fig:inversedemand_D}
\end{figure}

Given the parameters $j$, $\hat p$ and the pdf $f(x)$ characterizing the customers' system,  ${\cal D}$ determines  the {\em inverse demand function}
\begin{equation}
\label{eq:p^d}
p^d(\eta) = h + {\cal D}(j; \eta),
\end{equation}
which depends on both parameters $h$ and $j$. 
A graphical representation is given in the Appendix, Section \ref{sec_app:graphic}, and the resulting demand curves are illustrated on Fig.~\ref{fig:inversedemand_D}.

The stable equilibria of the demand satisfy 
\begin{equation}
 {\cal D}'(j; \eta)  =j-\Gamma'(\eta) \;\le 0
\label{eq:demandstab}
\end{equation}
where the prime, $'$,   
denotes the derivative with respect to $\eta$ at $j$ fixed. Equilibria with ${\cal D}'>0$ are unstable and are usually excluded in economic discussions.

From the stability condition (\ref{eq:demandstab}), 
and noting that $\Gamma'$ has a unique absolute minimum  
$\Gamma'(\eta_B)=\min_{\eta} \Gamma'(\eta) = \frac{1}{f_B} \; > 0$, where $f_B\equiv\max_x f(x)$, 
one gets that there is a critical value $j_B$ of the social strength, defined by
$j_B\equiv  1/f_B$:
whenever $j\leq j_B$ there is a unique solution
$\eta^d(\hat p,j)$ to (6), whereas
for $j > j_B$ there is a range of $\hat p$ values with  multiple solutions. Indeed, for $j>j_B$,
${\cal D}$ has two relative extrema\footnote{For multi-modal pdfs, the maximum number of such 
extrema is $1$ plus the number of modes of the pdf \cite{GoNaPhSeM3as}. Note that in the context of the RFIM, most studies consider the unimodal, Gaussian, case, or the bimodal case with $H_i=\pm H$.}, a minimum at $\eta_L(j)$ and a maximum at $\eta_U(j)$, with $\eta_L(j) \leq \eta_B \leq \eta_U(j)$ (see the Appendix, Fig \ref{fig_app:inversedemand_D}). These limiting values are given by the marginal stability condition 
\begin{equation}
 {\cal D}'(j;\eta)=0
\label{eq:dD=0}
\end{equation}
leading to
$\Gamma'(\eta_U) = \Gamma'(\eta_L)=j$, and the corresponding $\hat p$ values, $\hat p_{\Lambda}(j) \equiv {\cal D}(j; \eta_{\Lambda}(j))$, $\Lambda=L,U$. One has $\hat p_B  \leq \hat p_L \leq \hat p_U$. 

The definition (\ref{eq:D}) of ${\cal D}$ and the condition (\ref{eq:dD=0}) imply that the phase boundaries are given by the Legendre transform of the 
function $\Gamma$ defined in (\ref{eq:Gamma}). This fact allows to get various generic properties of the phase diagram, resulting from convexity properties that are associated to the Legendre transform (see \cite{GoNaPhSeM3as} for details). These properties have not been studied before in the context of the RFIM, where the focus is rather on the nature of the transitions, with  universal properties which are known, or assumed to be, independent of the distribution of the random field.

\subsection{Customers'phase diagram}
\label{sec:cpd}
We summarize the properties of the demand on a phase diagram of abscissas $j$ and ordinates $\hat p$, where we represent the boundaries between phases of qualitatively different behaviors. Whenever $\hat p$ and/or $j$ change across one of these boundaries, a stable demand equilibrium appears or disappears abruptly. Thus, these boundaries are lines of non-analyticity.

\begin{figure}
\centering
\includegraphics[width=0.6\textwidth]{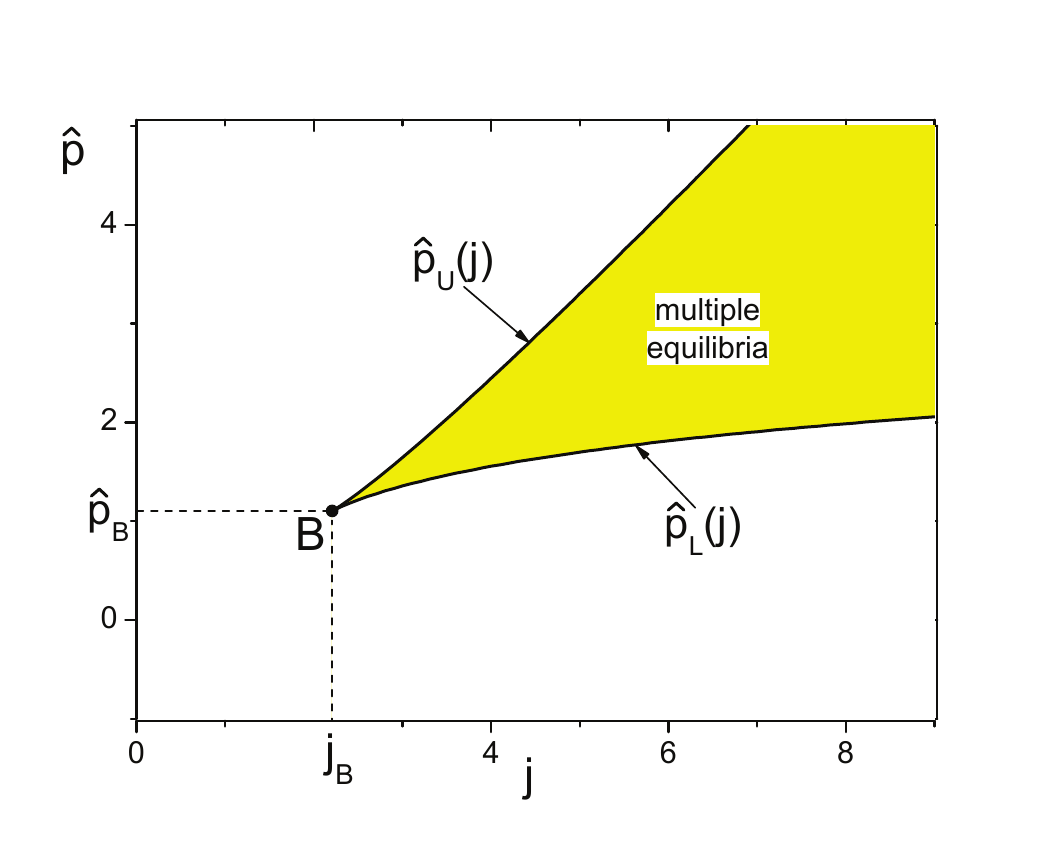}
\caption[]{\em Customers' phase diagram in the plane $\{ j$ (normalized social strength), $\hat p = p-h$ (posted price reduced by the mean willingness to pay in the population)$\}$, 
 for a smooth unimodal IWP distribution, illustrated with the case of the logistic distribution. The curves $\hat p=\hat p_L(j)$ and $\hat p=\hat p_U(j)$ delineates the domain of coexistence of two equilibria (one with a high demand and one with a low demand), and meet at the critical point $B$.}
\label{fig:customers}
\end{figure}

Figure \ref{fig:customers} presents the phase diagram corresponding to the unitary variance logistic distribution. 
The multiple solutions region (shadowed on the figure) appears at the bifurcation point $B \equiv (j_B,\hat p_B)$. 
The boundaries are given by the curves  $\hat p=\hat p_L(j)$ and $\hat p=\hat p_U(j)$ where $\hat p_L$ is concave and $\hat p_U$ is convex. 
At the bifurcation point $B$ the two branches meet forming a cusp, 
and beyond $j_B$ the width of the multiple solutions region increases 
monotonically with $j$, as illustrated in the figure. 

Whenever $\hat p_L(j) < \hat p < \hat p_U(j)$, 
there are thus multiple possible demands given by the two stable solutions,  
that we denote $\eta^d_L(j,\hat p)$ on the low-$\eta$ manifold, with $0 < \eta^d_L(j,\hat p) < \eta_L(j)$, and $\eta^d_U(j,\hat p)$ on the large-$\eta$ manifold, with $\eta_U(j) < \eta^d_U(j,\hat p) < 1$. There is a gap in the possible (stable) equilibrium values of $\eta$: the interval $[\eta_L,\eta_U]$ corresponds to unstable equilibria. The demand equilibrium in the large-$\eta$ manifold, $\eta^d_U$, is Pareto optimal: not only it gives positive surpluses to more individuals, but these surpluses are systematically larger than those corresponding to the lower demand equilibrium $\eta^d_L$, because the surplus function (\ref{eq:u_i}) increases with $\eta$. The high demand (Pareto optimal) equilibrium, $\eta=\eta^d_U$, requires customers coordination: it is the demand sustained by the social interactions. If coordination fails, the system stabilizes in the low demand 
equilibrium, $\eta=\eta^d_L$, on which everybody has a smaller surplus. 
As we will see in the next Section,  
the possibility of coordination failures has dramatic consequences on the pricing strategies.

\section{The Supply: Monopoly Market}
\label{sec:monopolist}
In this section we examine the consequences of customers social interactions on 
the pricing strategy of a monopolist who has to determine the price $P$ and 
how many units $N \eta$ of the good to put in the market in order to maximize his profit. For simplicity we assume a constant cost per unit of produced good. We aim at analysing the optimal strategies of the seller given the full knowledge of the demand structure discussed above, considering the markets equilibria (where offer equals demand), 
 with no strategic behaviour from the customers - they do not make anticipations on the seller's pricing strategy: for each posted price, the demand is given by the Nash equilibria discussed in the previous section. 
The results will then allow to discuss cases where the seller would not have all the detailed informations on the customers characteristics, and cases of 'iterated game', where customers buy or not at each instant of time (as going or not to a given popular restaurant every week end).

\subsection{Hidden market structure}
A standard economics viewpoint is that there exists, on one side,  a Demand curve, relating the price  to the quantity that consumers are willing to buy at that given price; and on the other side, a Supply curve, giving the price that the seller will post in function of the quantity asked by the consumers. The Demand curve (resp. the Supply curve) is  assumed to be monotonically decreasing (resp. increasing) with increasing quantity, and the market (or equilibrium) price is given by the intersection of the two curves. As recalled in the previous section, 
in the present case of market with externalities (the fact that the willingness to pay is increased by a social influence term), 
there is a wide range of parameters for which the Demand curve is not monotonic, and only the parts with negative slope can correspond to stable equilibria: for $j>j_B$ the domain of accessible $\eta$-values is disconnected. On the Supply side, in the present model there is no Supply curve {\it per se}: it is here assumed that the seller can produce any quantity of goods, and he/she will produce the quantity corresponding to the price that maximizes his/her profit. As detailed in the Appendix, Section \ref{sec_app:effectivesupply}, this optimization program has an interesting underlying mathematical structure for several reasons. First, as presented below Section \ref{sec:effectivesupply}, it leads to an effective Supply curve: the maximization of the profit can be written as the solution of an equality between the demand (as computed in the previous section) and an effective supply function. This allows for a discussion within an (almost) classical economics setting (in Physics one would say that this 
gives a physical interpretation of the underlying market 
structure).  
Second, there is an unexpected formal analogy between the demand problem and the profit maximization problem. As shown Section \ref{sec:analogy} below, the maximization of the profit leads to a system of equations which has the very same mathematical structure are the one for the demand shortly presented Section \ref{sec:demand}. This allows us to simplify the analysis of the supply problem. To our knowledge, these properties have not been noted before in the economics literature.  Finally, despite the simplification brought by the analogy, the pricing issue remains more involved than the
analysis of the demand, since for $j>j_B$ there is a range of prices for which the demand is multi-valued, and the domain in $\eta$ where the profit has to be maximized is non-connected.  

We first briefly present the mathematical analysis of the profit maximization problem that has to be solved by the seller, as evoked above.  
We then discuss the optimal pricing strategies as a function of $j$ and $h$, the parameters characterizing the customers population. Most mathematical details are left to the Appendix. 

\subsection{Supply function} 
\label{sec:maxpi}
\subsubsection{The optimization problem}
 If $C$ is the production cost per unit, the seller's profit is $\Pi=  (P-C) \; N\eta$. 
For simplicity we assume that $C$ is independent of the total number of produced units. One can easily find many examples of constant marginal cost, and in some cases quasi-null marginal cost, like the cost of replicating software, musical or audiovisual digital files\footnote{One can note that a standard decreasing marginal cost would lead, in the large $N$ limit, to a total cost that can be neglected (compared to $N\eta$). Taking into account a total cost that depends on $N \eta$ would require an analysis of finite size effects, which is out of the scope of the present paper.}.
\begin{figure}
\centering
\includegraphics[width=0.60\textwidth]{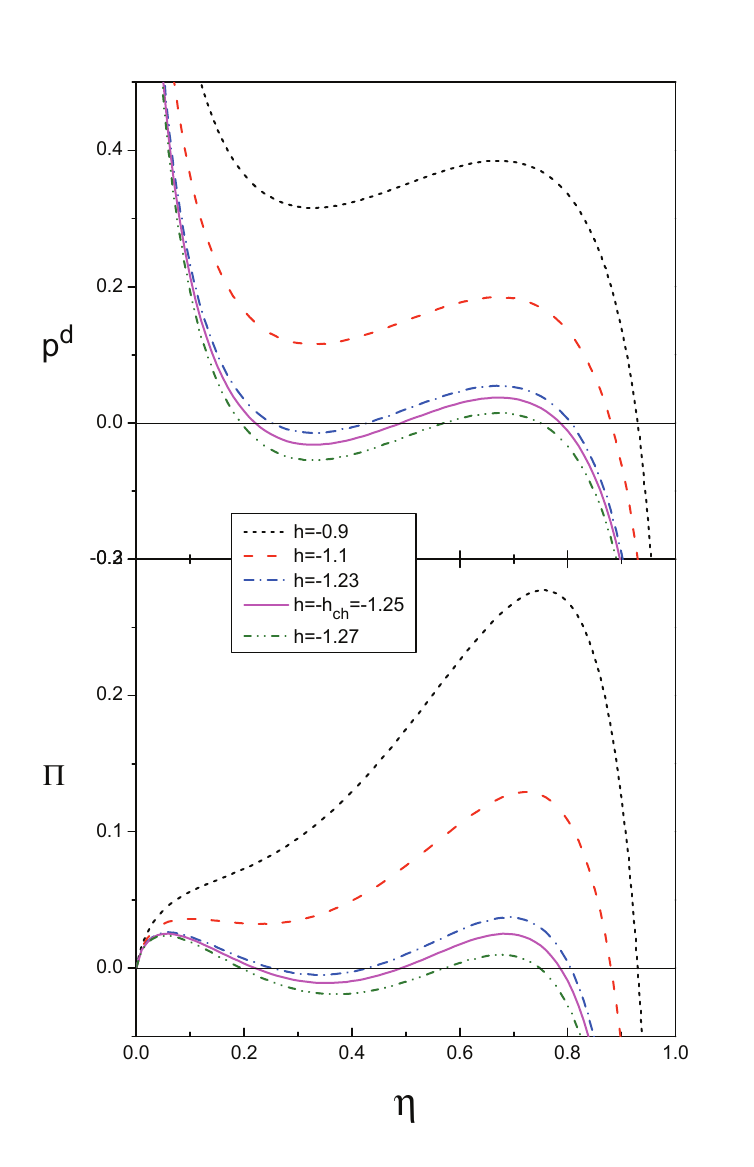}
\caption[]{\em Price (upper figure) and profit (lower figure) for different values of the mean willingness to pay $h$, and for the social strength $j=2.5$. Upon increasing $h$ from large negative values, at $h=h_{ch}$ the optimal seller's strategy (the one giving the largest profit) 
changes from a high price-low supply strategy 
to a low price-high supply strategy.} 
\label{fig:price-profit}
\end{figure}

The relevant normalized variables corresponding to the supply are 
$\frac{H-C}{\sigma}$ and $\frac{P-C}{\sigma}$.   
Since the demand only depends on the difference $\hat p=p-h=\frac{P-H}{\sigma}$, which is invariant under  
the shift of $C/\sigma$ on both $p$ and $h$, we can use the results of the demand analysis without any change. Thus, hereafter we interpret $h$ and $p$ as excess values with respect to the normalized cost $C/\sigma$. With this convention, the monopolist has to choose the price $p$ in order to maximize his 
normalized (in the same sense as equation (\ref{eq:reduced_var})) profit
\begin{equation}
\label{eq:profit2}
\pi(\eta,p) \equiv \frac{\Pi}{N\,\sigma} = p \; \eta, 
\end{equation}
under the assumption that $\eta$ and $p$ are related through the demand equilibrium equation (\ref{eq:p^d}). The maximum of the profit may correspond to a low demand with a high price or to a high demand and a low price, depending on the customers characteristics (mean willingness to pay and strength of the social interactions) in a non trivial way, as may be seen on Figure \ref{fig:price-profit}. 

The optimal price is given by the solution(s) of $\partial \pi / \partial p=0$ that satisfy the maximum (second order) condition $\partial^2 \pi / \partial p^2 < 0$, and by inspection of the profit at the boundaries of the accessible $\eta$ domain.
\subsubsection{Analogy with the demand analysis}
\label{sec:analogy}
The maximization problem can be cast into the same mathematical framework as the analysis of the demand.
That is, the extremum condition $\partial \pi / \partial p =0$ can be written 
as the following condition on the quantity $\eta$ that has to be supplied by the monopolist, 
\begin{equation}
-h =  \widetilde{\cal D}(j; \eta)
\label{eq:h_monop1}
\end{equation}
and the second order condition $\partial^2 \pi / \partial p^2 < 0$ can be written as
\begin{equation}
 \widetilde{\cal D}'(j; \eta) \le 0,
 \label{eq:max_cond2} 
 \end{equation} 
where 
$\widetilde{\cal D}(j; \eta)  \equiv \frac{ d }{d \eta } [ \eta \;{\cal D}(j; \eta) ] $  
is 
\begin{equation}
 \widetilde{\cal D}(j; \eta) = 2 j \eta - \widetilde{\Gamma}(\eta)
\label{eq:Dtildbis}
\end{equation}
with the function $\widetilde{\Gamma}$ is defined by
\begin{equation}
\widetilde{\Gamma}(\eta) \equiv \frac{ d }{d \eta } [ \eta \Gamma(\eta) ]. 
\label{eq:hatgamma} 
\end{equation}
Comparing (\ref{eq:Dtildbis}) with the definition of the demand function ${\cal D}$, Equ. (\ref{eq:D}), and (\ref{eq:h_monop1}) and (\ref{eq:max_cond2}) with the equations for the Demand, (\ref{eq:p-h})  and (\ref{eq:demandstab}) respectively, one sees that the 
supply equations have the very same structure as the demand ones, with $-h$, $\widetilde{\Gamma}$, $\widetilde{\cal D}$ and  $2 j$ playing the roles of, respectively, $\hat p=p-h$, $\Gamma$, ${\cal D}$ and $j$. In order to take full advantage of this analogy, we need to assume hereafter that, like $\Gamma(\eta)$, $\widetilde{\Gamma}(\eta)$ is monotonically increasing from $-\infty$ to $+\infty$ when $\eta$ goes from $0$ to $1$, and that it has a single inflexion point. 
As detailed in the Appendix, Section \ref{sec:app_Gamma_hat}, this is true 
under a not very stringent condition on the IWP distribution.
 
Equation (\ref{eq:h_monop1}) and (\ref{eq:max_cond2}) characterize the {\em relative} maxima of the profit in interior regions of the accessible $\eta$-domain. However, as already said, for $j>j_B$ and $\hat p_L(j) < \hat p < \hat p_U(j)$, values of $\eta$ in the range $[\eta_L(j),\eta_U(j)]$ should not be considered as they correspond to unstable equilibria. As a consequence, 
there may exist maxima at the boundaries of the forbidden interval. In Section \ref{subsec:LowEtaBranch} we show that when they exist, these  are always sub-optimal maxima so that we may first ignore them in the 
following. Nevertheless, as we will see in 
Section \ref{sec:discussion}, they play an important role in the discussion of the possible monopolist's strategies.

\begin{figure}
\centering
\includegraphics[width=0.80\textwidth]{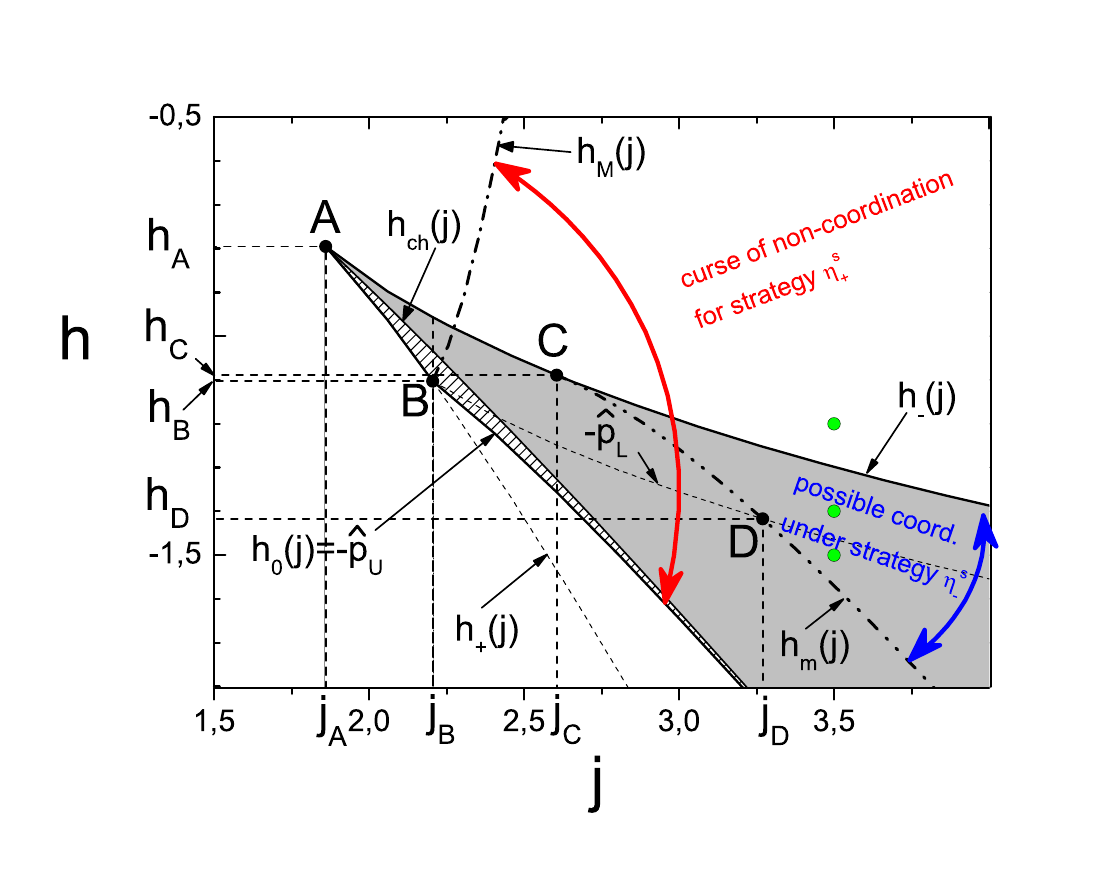}
\caption[]{\em Supply phase diagram. For each point in the plane $(j,h)$ there is one or several values of the price $p$ giving a (possibly relative) maximum of the profit. A pricing strategy at given values of $(j,h)$ is here the choice of which price to post. The different curves delineate domains with one or several maxima of the profit, and with different qualitative properties, notably with respect to the customers equlibria. The main features of the phase diagram are as follows.\\
(1) Seller's dilemma (see text, section \ref{sec:dilemma}): for $h_+ \le h \le h_-$ there are multiple sub-optimal pricing strategies. The line $h_{ch}$ separates the region $h>h_{ch}$ where the optimum is to sell to large fractions of customers at low prices, from the region with $h<h_{ch}$ where the optimum is to sell to a small fraction of customers at relatively high prices. The latter strategy, in the domain $h_+ \le h < h_{ch}$ where it is suboptimal, is in fact only viable above the curve $h_0(j)$ on which the associated profit vanishes. 
\\(2) Large domain of systemic risk (the coordination curse, see text, section \ref{subsec:HighEtaBranch}): within the region below $h_M(j)$ (large heavy double-arrow line), at the price targetting the largest profit the demand is multivalued. Hence the high-demand strategy (targetting $\eta=\eta^s_+$) may fail due to lack of customers coordination.\\
(3) For $-\hat p_L(j) < h < h_-(j)$ there exists a third sub-optimal strategy (never optimal), which corresponds to targeting the maximum possible fraction of customers on the low demand equilibrium. 
\\(4) The small (rightmost) heavy double-arrow line indicates the region $-\hat p_m(j) < h < h_-(j)$ and $j>j_C$ where a seller targeting the low-demand equilibrium ($\eta=\eta^s_-$), might benefit from a coordination success. \\
Values for the logistic distribution: $j_A=27 \sqrt{3} / (8 \pi) \approx 1.86, h_A \approx -0.80 $, $j_B \approx 2.21, h_B \approx -1.10$, $j_C \approx 2.61, h_C \approx -1.09$, $j_D \approx  3.27, h_D \approx -1.42$. The three green dots on the figure correspond to values of $(j,h)$ used on Figure \protect{\ref{fig:examplesj35_h-12_h-14_h-15}} (see the Appendix).}
\label{fig:MonopolistCompletePhD}
\end{figure}
\subsubsection{An effective supply function}
\label{sec:effectivesupply}
With these clarifications, the analysis follows the same lines as that of the  demand. Given the parameters $h$ and $j$ characteristic of the customers population, the solution(s) of $-h =\widetilde{\cal D}(j; \eta)$ -- eq. (\ref{eq:h_monop1}) -- determine $\eta^s \equiv \eta(j,h)$ which gives the number of goods $N \eta^s$ to be supplied by the monopolist. Introduction of $\eta^s$ into  (\ref{eq:p^d}) allows to determine the price $p^s=p^d(\eta^s)$ to be posted which, interestingly, is given by the difference between the 
function ${\cal D}$ 
of the demand and the  
function $\widetilde{\cal D}$ of the supply optimization: 
\begin{equation}
\label{eq:p^s}
p^s(j,h)={\cal D}(j; \eta^s)- \widetilde{\cal D}(j; \eta^s).
\end{equation}

Alternatively, the solution of seller's problem  can be expressed as finding the solution(s) $\eta^s$ of the equality between demand and supply:
\begin{equation}
\label{eq:p_monop_main}
p^d(\eta)=p^s(\eta)
\end{equation}
where $p^s(\eta)$ is an effective supply function, defined by
\begin{equation}
p^s(\eta) \equiv -\eta {\cal D}'(j;\eta) = \eta [\Gamma'(\eta)-j]
\label{eq:ps2_main}
\end{equation}
(see the Appendix \ref{sec_app:effectivesupply} for details).

The corresponding profit is obtained through introduction of $\eta^s$ and $p^s$ 
into (\ref{eq:profit2}). If several solutions exist, the monopolist should select the one corresponding to 
the maximum maximorum of the profit. 

\section{The Monopoly Phase Diagram}
\label{sec:monopolist_phase_diag}
 In this section we discuss the optimal pricing strategies as a function of the parameters characterizing the customers population. Like for the demand, we represent the the boundaries of regions with similar properties on a phase diagram (of axis $j$ and $h$ in this case). Whenever  $h$ and/or $j$ change across any of these boundaries, 
either the optimal price jumps discontinuously or the number of pricing strategies (extrema of $\pi$)  changes. Taking advantage of the formal analogy pointed out in the previous section, the construction of the phase diagram follows the same lines as for the demand. However, the supply problem is more involved because for $j>j_B$ 
one has to take care of the fact that for some range of parameters there are two possible demands, and in that case there is a range of unstable demand equilibria, $\eta \in [\eta_L(j),\eta_U(j)]$. 
 Figure \ref{fig:MonopolistCompletePhD} shows a complete phase diagram for the particular case of a logistic distribution. Its main features, that we discuss 
thoroughly in the following sections, are generic for any smooth distribution of the IWP having a single maximum and satisfying condition (\ref{eq_app:cond_on_f_b}). Details of the construction are given in the Appendix, Section \ref{sec:app_phasediag}.

\subsection{Seller's Dilemma} 
\label{sec:dilemma}
In this section we discuss the domain of existence of multiple strategies,  
in which the seller 
faces the following dilemma: should he sell at low price to a large number of customers, or at high price to a small number of customers?
\subsubsection{A first order transition}
Like for the demand, there is a bifurcation in the behavior of the optimal price, 
that here takes place at a critical value of $j$ defined by
\begin{equation}
j_A=\widetilde{\Gamma}'(\eta_A)/2, \;\;\;\; {\rm with} \;\;\;\;\eta_A \equiv \arg \min_{\eta} \widetilde{\Gamma}'(\eta).
\label{eq:j_A}
\end{equation} 
with $\widetilde{\Gamma}$ defined in  (\ref{eq:hatgamma}). This gives in the plane $(j,h)$ a critical point $A=(j_A,h_A)$, analogous for the supply to the point $B$ in the demand phase diagram. One can show that  
$\eta_A  \leq  \eta_B$ and $j_A  \leq  j_B$. 

When $j>j_A$ the extremum condition for the profit, equation (\ref{eq:h_monop1}), presents multiple solutions for some range of values of $h$. 
In the $(j,h)$ plane, the boundaries of the domain of multiple solutions are given by the Legendre transform of the function $\widetilde{\Gamma}$.
More precisely, they are given by the curves $h_\pm(j)=-\widetilde{\cal D}(j; \eta_{\pm}(j))$ where $\eta_-(j) < \eta_A$ and $\eta_+(j) > \eta_A$ are, respectively, the minimum and the maximum of $\widetilde{\cal D}(j; \eta)$, both satisfying the marginal stability condition $\widetilde{\cal D}'(j,\eta_\pm)=0$
(for an illustration, see the left-hand side panels of Figure \ref{fig:examplesj35_h-12_h-14_h-15}, in section \ref{sec:Numerical_illustrations}). 
Note that these curves $h_\pm(j)$  are the analogous, in the monopolist pricing problem, 
of the curves $\hat p_{U,L}(j)$  
giving the boundaries of the multiple solution domain for the demand, as discussed Section \ref{sec:customers}.

If $(j,h)$ lies outside the multiple solutions region, the profit 
as a function of $\eta$ has a single maximum. 
Thus, the profit optimization has a unique solution, and the optimal supply $\eta^s$, the price to be posted $p^s$ and the expected profit are obtained   
using equations (\ref{eq:h_monop1}), (\ref{eq:p^d}) and (\ref{eq:profit2}) respectively. 

\begin{figure}
\centering
\begin{tabular*}{\textwidth}{cc}   
\includegraphics[width=0.48\textwidth]{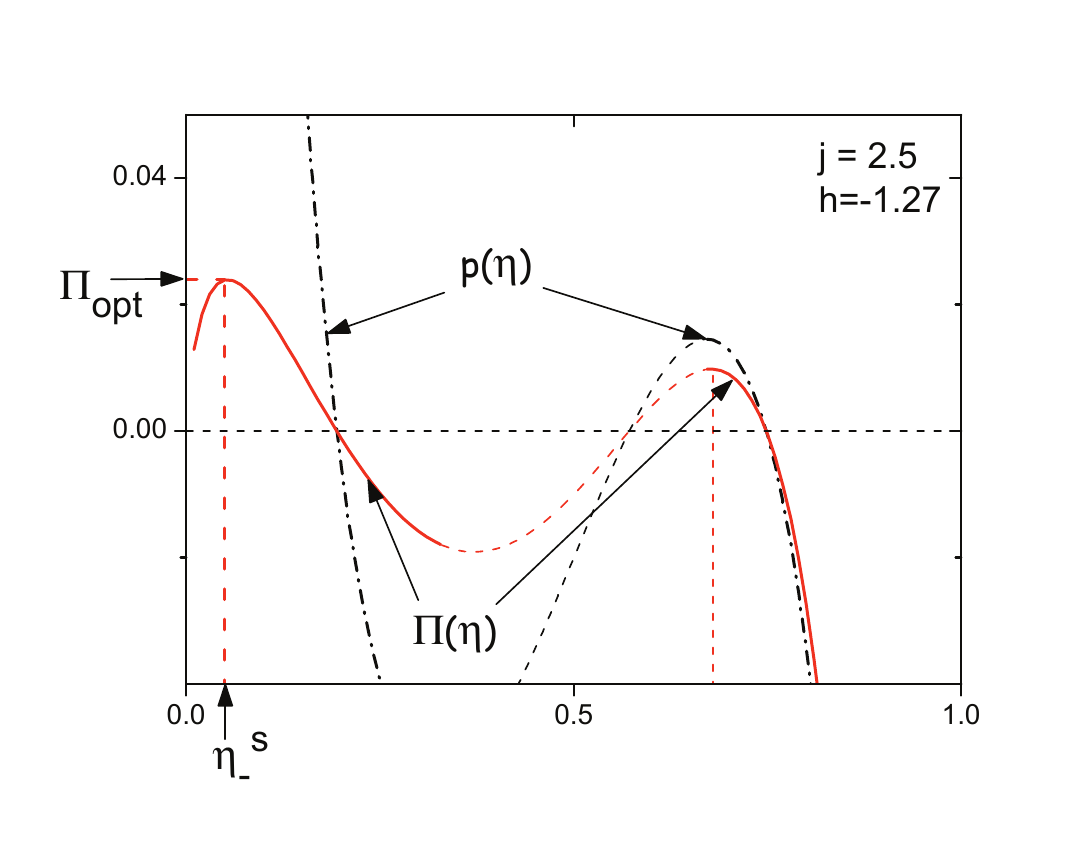} & \includegraphics[width=0.50\textwidth]{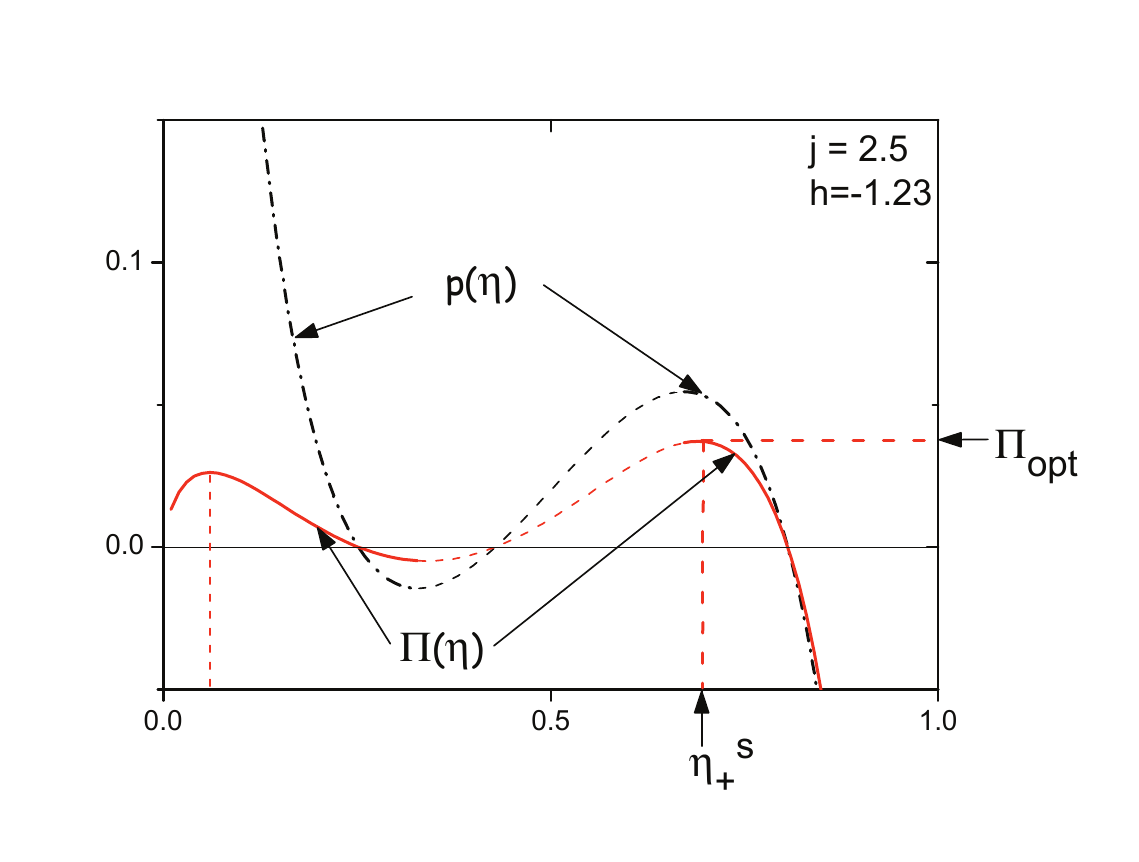} \\
(a) \& (b)\\
\end{tabular*}
\caption[]{\em Price and profit as a function of the supply $\eta$ for $j=2.5$ (for which $h_{ch} \approx -1.247$), for two values of $h$. The supplies $\eta^s_-$ and $\eta^s_+$ and the profits of the optimal strategies are indicated. (a) $h=-1.27 < h_{ch}$: the optimal strategy lies on the low-$\eta$ branch ; (b) $h=-1.23 > h_{ch}$: the optimal strategy lies on the high-$\eta$ branch, at a value of $\eta$ very close to the point where the high-$\eta$ branch disappears. }
\label{fig:pPij25h-127-123}
\end{figure}

Within the domain of multiple solutions (hence  $j>j_A$ and $h_+(j)< h <h_-(j)$), the two local maxima of the profit correspond to two different strategies for the monopolist: either to attract a large fraction of buyers at low prices (solution $\eta^s_+(j,h)$) or to target few buyers willing to pay high prices (solution $\eta^s_-(j,h)$). There is a first order transition at a critical value $h_{ch}(j)$ within the coexistence region, where the optimal strategy (corresponding to the maximum maximorum of the profit) switches abruptly from one branch to the other. On this line $(j,h_{ch}(j))$, both strategies give the same profit (and this is what defines $h_{ch}(j)$), but the corresponding prices are very different. Figure \ref{fig:pPij25h-127-123} illustrates the transition between the low-demand and the high-demand optimal strategies when $h$ changes across the value $h_{ch}(j)$ for $j=2.5$. Another example of this strategy-switching has been presented in \cite{GoNaPhVa05} for a particular IWP distribution. 

At fixed $j  >j_A$, if $h$ is small enough to be outside the coexistence region, the optimal 
price is unique. 
If $h$ increases, this maximum of the profit can be followed into the coexistence region, where it corresponds to values $\eta^s_-(j,h)$ on the low-$\eta$ branch of $\widetilde{\cal D}$, up to $h=h_-(j)$ where it disappears. Conversely, decreasing $h$  
from a value high enough to have a unique demand, the solution $\eta^s_+(j,h)$ lying on the large-$\eta$ branch of $\widetilde{\cal D}$ is met upon entering the coexistence region and can be followed down to $h=h_+(j)$ where it disappears.

\subsubsection{Intertwining between the supply first order transition and the demand multiple equilibria}
Besides the point $A$, there is an additional singular point, $B = \{j_B, h_B = -\hat p_B \}$, a mirror of the customers' point $B$, lying on the boundary line $h=h_+(j)$. For $j> j_B$, the domain of existence of the high-supply strategy $\eta^s_+$ has a lower bound of viability given by the condition that the optimal price for the high-$\eta$ strategy has to be positive. There is a line where the optimal price for the high-$\eta$ strategy is zero ---that is, given our convention, the actual 
optimal
price is  equal to the production cost. This null-price curve is given by $h_0(j) \equiv -\hat p_U(j)$, where $\hat p_U(j)$ is the boundary of the customers 
$U$ solution (see equation (\ref{eq:dD=0})).  
As may be seen on Figure \ref{fig:MonopolistCompletePhD} 
the line $h_0(j)$ starts at point $B$ on the line $h_+(j)$ and verifies $h_0(j) > h_+(j)$ for all $j>j_B$.
 
Interestingly, within the range $j_A < j < j_B$, the monopolist has two possible strategies but the demand is single valued. Hence, 
the seller can drive the customers to the optimal profit equilibrium by just posting the corresponding price.

In contrast, for $j>j_B \, (\ge j_A)$, as seen in Section \ref{sec:demand_function} the customers system has multiple  equilibria: a low-$\eta$ risk-dominant one and a high-$\eta$ Pareto-optimal equilibrium which needs coordination of the buyers. The actual equilibrium reached by the customers system may not be the one targeted by the monopolist at the posted price, due to coordination failure. This problem has been mentioned by Becker in his note on restaurant pricing strategies \cite{Becker91}. 
In the next section we determine precisely the parameter region in the phase diagram where choosing a pricing strategy can be subject to such epistemic uncertainty.

\subsection{Pricing when the demand is multi-valued} 
\label{sec:risk}
In this section we are interested in the possible outcomes when the values of $h$ and $j>j_B$
are such that the optimal price for the seller may correspond to a situation where the demand is multi-valued. 
Then, when posting a low price in order to meet the optimal large demand, the seller 
runs the risk of finding 
himself with a low demand if coordination of the customers fails.

Consider the demand $\eta=\eta^d$ as function of the price $p$. As discussed in section \ref{sec:customers}, within the range of interest here the demand has two branches, one with low $\eta$ ($0 \leq \eta\leq \eta_L$) and another with high-$\eta$ ($\eta_L < \eta_U \leq \eta\leq 1$ ). These exist for a range of (shifted) prices $[\hat p_L(j), \hat p_U(j)]$, where the two branches overlap. We can look at the profit as one increases the price, following separately each branch. 
The profit may have (relative) maxima at some of the boundaries ($\eta=0, \eta_L(j), \eta_U(j), 1$), and/or in the interior parts, the latter giving the solutions $\eta^s_-$ on the low-$\eta$ branch and  $\eta^s_+$ on the large-$\eta$ branch (see Section \ref{sec:app_lowlowhighhigh} for more details). 

Whenever the monopoly chooses the low-supply strategy $\eta^s_-$, it expects that the customers system selects the low-$\eta^d$ solution. Conversely, if the monopoly chooses the high-demand strategy $\eta^s_+$, it expects  
the customers system to select  
the large-$\eta^d$ solution. 
Thus, we can first analyze each branch separately. 
Numerical illustrations of the different situations are given in the Appendix, 
Section \ref{sec:Numerical_illustrations}.
\begin{figure}[htbp]
\centering
\includegraphics[width=0.75\textwidth]{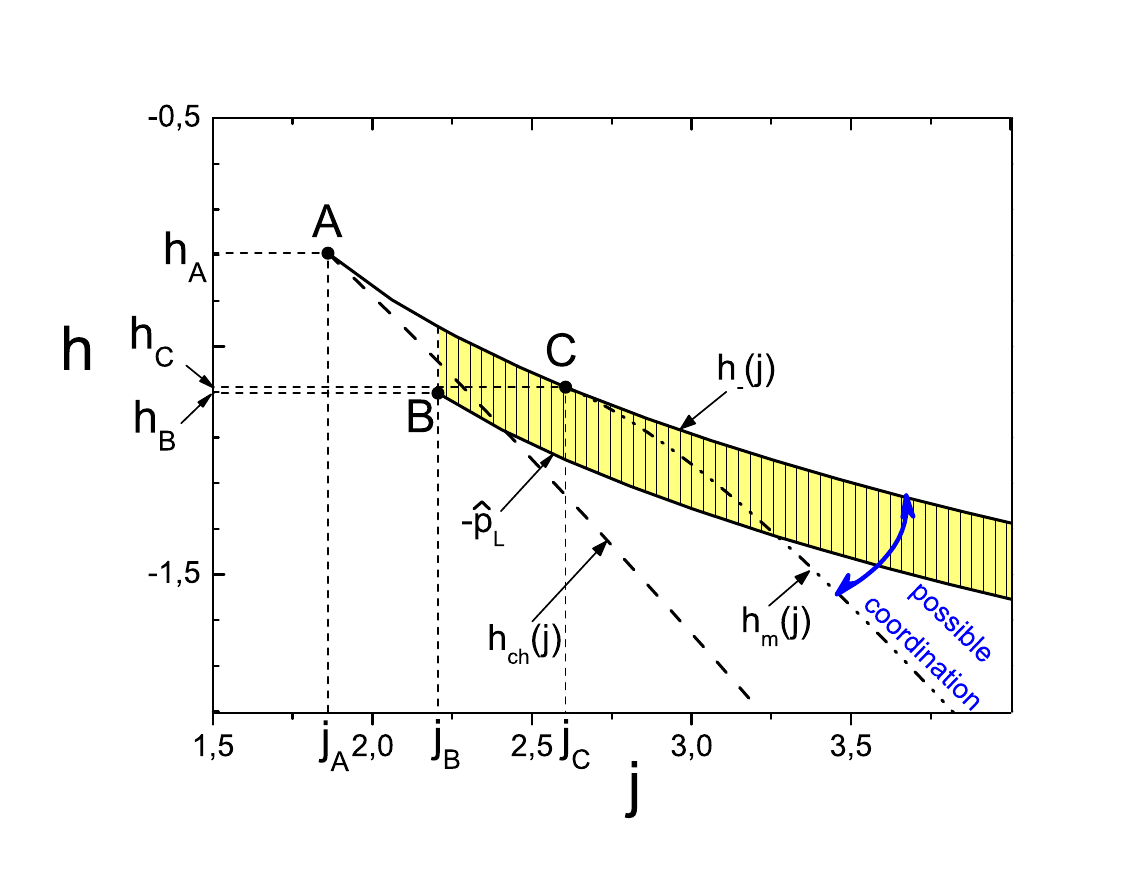}
\caption[]{\em Supply phase diagram: properties of the low-$\eta^s$ extrema, that exist for any $h<h_-(j)$ and $j>j_A$ (for $j<j_A$ there is a single optimal supply). The low-$\eta^s$ solution corresponds to the absolute maximum of the profit whenever $h<h_{ch}(j)$. For $j>j_B$ and for values of $h$ inside the 
hatched region ($- \hat p_L(j)<h<h_-(j)$) there is, besides the maximum at $\eta^s_-$, a maximum of $\pi$ at $\eta_L$ for a price $p=\hat p_L(j) +h$. In the region $h_m(j)<h<h_-(j)$ 
a (not optimal) strategy targeting $\eta^s_-$ may give raise to coordination among customers. In that case, the fraction of buyers, hence the profit, is larger than expected. }
\label{fig:MonopolistPhD_LowEta}
\end{figure}						

\subsubsection{Pricing without 
need of coordination - Low-demand manifold}
\label{subsec:LowEtaBranch}
Consider first the (relative) maximum(s) of the profit at low values of $\eta$, i.e. in the domain $\eta \in [0,\eta_L]$. These solutions correspond to fractions of buyers on the low-$\eta^d$ branch of the demand 
where the customers' equilibrium does not need coordination. We have seen that it corresponds to the optimal strategy below the first order transition, that is whenever $h < h_{ch}(j)$ (see Figure \ref{fig:MonopolistPhD_LowEta} for the details).

On the low-$\eta^s$ branch ($h < h_- $) the profit has one maximum at $\eta^s_-$, and, for $-\hat p_L < h < h_- $, a second maximum at the boundary $\eta_L$, which is always suboptimal (see the Appendix, section \ref{sec:app_nonoptetaL}).

The solution $\eta^s_-$ may lie in the region where the customers' system has multiple equilibria. We find (Appendix, Section \ref{sec:app_loweta}) that this happens above the line $h_m(j)$, a region inside the coexistence domain 
(see Figures \ref{fig:MonopolistCompletePhD} and \ref{fig:MonopolistPhD_LowEta}, heavy double-arrow line)   
where this  solution $\eta^s_-$ exists but is not the best one. 

Suppose that, starting from a value of $h$ below $h_{ch}(j)$, $h$ increases. 
Assuming smooth (``adiabatic'') changes in the population parameters and behaviour,
the posted price has to change smoothly, following the demand on the low $\eta$ branch. 
Up to $h=h_m(j)$, the posted price remains in the region for which $\hat p$ gives a unique solution for the demand.
Beyond $h_m(j)$,  that is for $h_m < h < h_-$ and $j$ larger than some value $ j_C$ (see Section \ref{sec:app_loweta}), the low-$\eta^s$ relative maximum is such that $\hat p_L < \hat p < \hat p_U$, i.e. it lies in a region where the customer system has multiple equilibria. 
Once in this regime, if for any reason the customers do coordinate, the demand and hence the resulting profit are much higher than the one expected without coordination. Note that the profit at this price 
is not the best profit that the seller can make when the demand lies on the high demand branch (see below).

One should note that no hint of the existence of an alternative strategy can be obtained from smooth changes in the price. Thus, if the seller lacks exact information on the population characteristics, he will remain unaware of the existence of the high-$\eta^s$ strategy.
Conversely, even if the seller knew that a high-$\eta^s$ solution exists, 
choosing the (suboptimal) local maximum on the low-$\eta$ branch represents a possible risk adverse strategy, allowing to avoid 
losses due to coordination failures.

\subsubsection{Pricing expecting coordination - The curse of coordination} 
\label{subsec:HighEtaBranch}
\begin{figure}[!ht]
\centering
\includegraphics[width=0.75\textwidth]{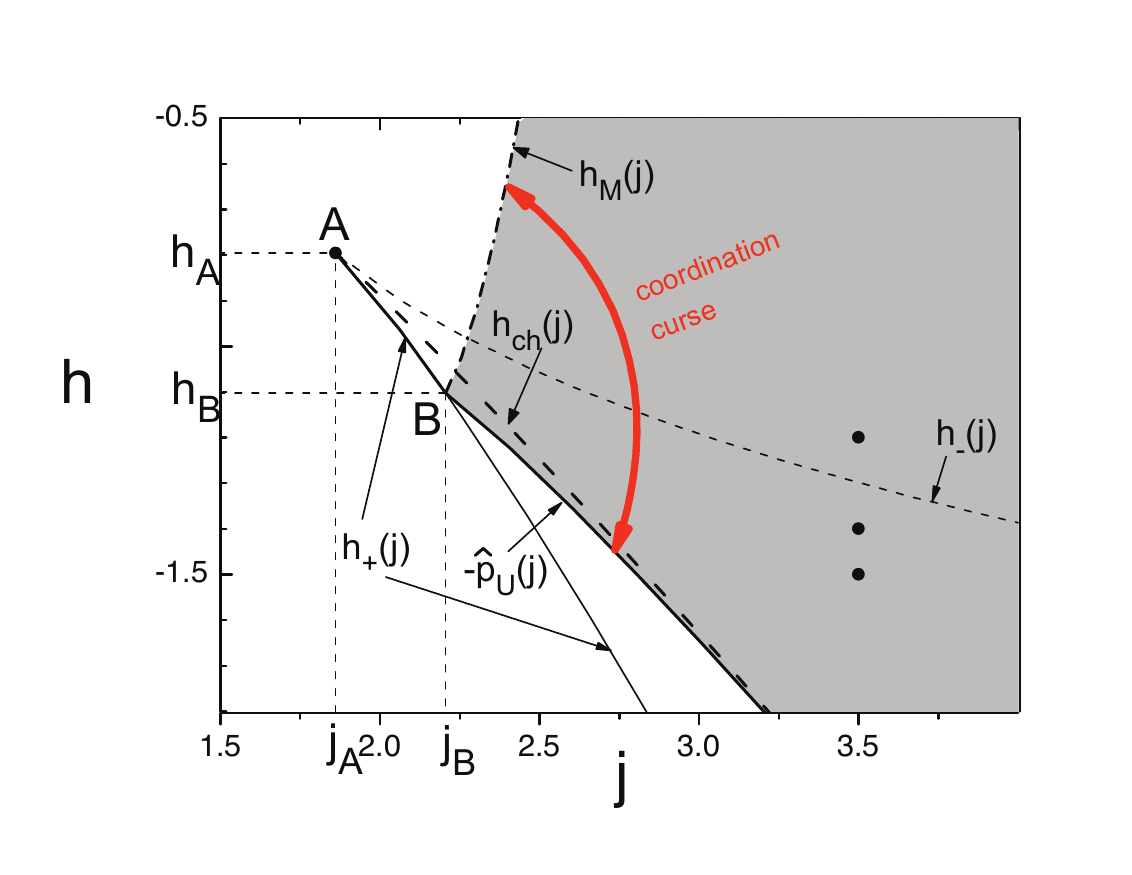}
\caption[]{\em Supply phase diagram: properties of the high-$\eta^s$ extrema that exist for any $h>h_+(j)$ if $j_A<j<j_B$, and for  $h>-\hat p_U(j)$ if $j>j_B$ (for $j<j_A$ there is a single optimal supply). The high-$\eta^s$  strategy targets a fraction of buyers $\eta^s_+ > \eta_+$. This solution corresponds to the absolute maximum of the profit whenever $h>h_{ch}(j)$. For $j>j_B$ and for values of $h$ inside the grey region ($- \hat p_U<h<h_M$), this strategy is optimal provided customers coordinate. If coordination is not achieved, the actual customers' equilibrium lies on the low demand 
branch, providing a profit smaller than expected. The three black dots correspond to values used on Figure \protect{\ref{fig:examplesj35_h-12_h-14_h-15}}.}
\label{fig:MonopolistPhD_HighEta}
\end{figure}

Let us now analyze the outcomes when the monopolist chooses to
target the large fraction of customers, $\eta=\eta^s_+$. 
As we have seen, for $j>j_B$ such solutions exist only for an average willingness to pay large enough ($h > h_0(j)$, see section \ref{sec:dilemma} and figures  \ref{fig:examplesj35_h-12_h-14_h-15}), with fractions of buyers $\eta^s_+$ on the high-$\eta$ branch 
of $\widetilde{\cal D}(j; \eta)$. They satisfy $\eta_U < \eta_+ \le \eta^s_+ \le 1$. We know that this strategy gives the largest profit above the first order transition line, that is for $h> h_{ch}(j)$.

Let us determine the range of $h$ values for which this strategy corresponds to  a price that falls within the domain where the demand is multi-valued. That is, we consider the possibility that the optimal price $p^s$ is such that 
$\hat p=p^s-h$ lies in the range $[\hat p_L(j), \hat p_U(j)]$, for which there are two possible demands, the one expected by the seller, and the one on the low $\eta^d$ branch. This is the case if $\eta^s_+$ lies in the range $[\eta_U, \eta_M(j)]$, where  $\eta_M(j)$ is the value on the high demand branch for $\hat p=\hat p_L(j)$. According to (\ref{eq:p-h}), $\eta_M$ is given by ${\cal D}(j;\eta_M)=\hat p_L(j)$.  
The limiting case where the seller's price $p^s$ corresponds to this demand $\eta_M$ is obtained when $h$ takes the value $h_M$ which satisfies the maximum condition (\ref{eq:h_monop1}) for $\eta=\eta_M$, that is
\begin{equation}
\label{eq:h_M}
h_M(j) \equiv - \widetilde{\cal D}(j;\eta_M) 
\end{equation}
The other limiting case corresponds to $\hat p=p^s-h=\hat p_L(j)$, which occurs if $h=- \hat p_U(j)$. 
Thus, for the mean willingness to pay in the range $[- \hat p_U(j), h_M(j)]$, the 
strategy $\eta^s_+$ drives the customers into the 
region of multiple solutions for the demand, making uncertain the actual outcome. In the plane $(j,h)$, the line $h_M(j)$  starts at the point $B$ (since $- h_M(j_B) = \widetilde{\cal D}(j_B;\eta_B) = \hat p_U(j_B)$), 
and increases with $j$ faster than the line $h=-\hat p_L(j)$ because for positive prices ${\cal D}'(j;\eta)<0$ 
so that (\ref{eq:h_M}) gives $h_M(j) > - \hat p_L(j)$.  
\begin{figure}[!ht]
\centering
\includegraphics[width=0.75\textwidth]{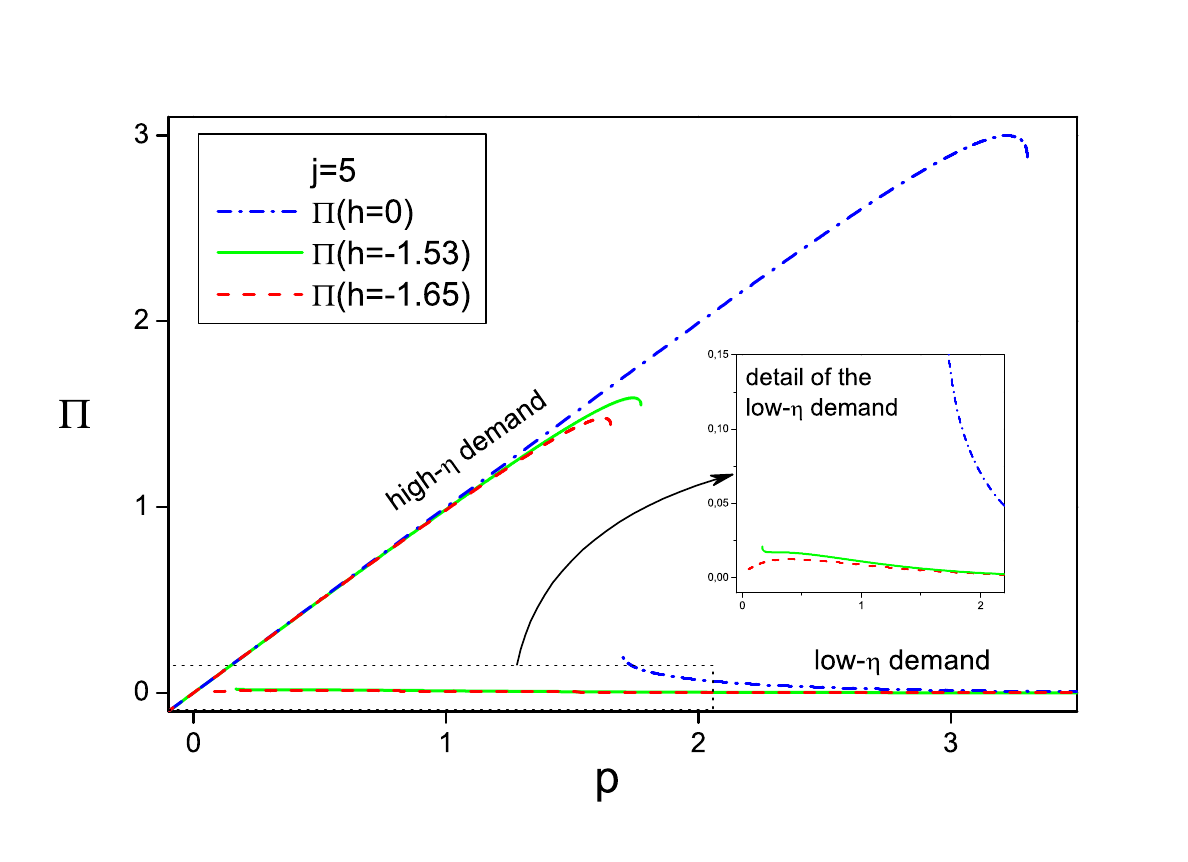}
\caption[]{\em Profit vs. price for $j=5$ and three different values of $h$. The optimal profit is obtained for the demand sustained by the social interactions (high $\eta$ branch), at a price slightly below the critical price at which the high demand disappears.}
\label{fig:pi_vs_p}
\end{figure}
Thus, posting a price corresponding to $\eta^s_+$ in the region where this strategy is optimal, i.e. for $h_{ch}(j) <h < h_M(j)$, may have an uncertain outcome because this price leads to the optimal profit only if the actual fraction of customers lies on the high-$\eta^d$ manifold. If the customers fail to coordinate, the fraction of buyers will correspond to the solution lying on the low-$\eta^d$ manifold at this same price, giving a profit much smaller than the expected one. The uncertainty region of the high-$\eta^s$ supply strategy for the case of the logistic distribution, represented  by a large heavy double-arrow line on figures \ref{fig:MonopolistCompletePhD} and \ref{fig:MonopolistPhD_HighEta}, is seen to cover a wide range of parameters, much larger than the region where multiple (locally optimal) strategies exist for the monopolist. 

In addition, another striking result is that 
the optimal strategy corresponds to a price value very close to the boundary  where the demand sustained by the social influence, that is the high-$\eta$ branch, disappears ($\eta^s$ slightly larger than $\eta_U$, $\hat p^s$ just below $\hat p_U$). This is illustrated on Figure \ref{fig:pi_vs_p} which shows the profit as a  function of the price, at a large value of $j$ and for different values of $h$ -- see also 
Figure \ref{fig:pPij25h-127-123} above and Figure \ref{fig:examplesj35_h-12_h-14_h-15}, Section \ref{sec:Numerical_illustrations}. Hence a small change in the population characteristics, if unexpected by the seller,  may 
drive the current posted price outside the domain of existence of the high profit solution. 

\section{Discussion: pricing under epistemic uncertainty}
\label{sec:discussion}
In his note on pricing of goods with bandwagon properties, like restaurants, best-seller books, successful plays in theatres, etc, Becker \cite{Becker91}  provides an explanation of apparently irrational pricing, that he attributes to the fact that the preferences for such goods are highly dependent on social influences. 
His qualitative analysis assumes implicitly that the optimal strategy for the seller corresponds to the large supply branch.
He argues that the price of these goods does not increase despite a chronic excess demand 
for fear of what he calls ``fickleness" of the consumers, that is nothing but the fragility of coordination. 
As we develop in this section, our mathematical modelling shows that there may be at least another explanation to the pricing behaviour. 

In the case of a restaurant, one can assume that there is essentially no strategic behaviour of the customers with respect to the seller, eventhough the customers 'play a repeated game': a same customer is likely to go several times to this same restaurant. One can thus assume a dynamical setting where at each instant of time the seller posts a (possibly different) price. A reasonable assumption is that for a small change in any parameter the demand will respond smoothly, whenever possible. Hence, if the system is in state $\eta^d$ on one branch of the multiple equilibrium region of the demand, away from the end points $\eta_L$ and $\eta_U$, a small change in price, i.e. in $\hat p$, will slightly shift the demand along the same branch.

\subsection{Possible strategies to overcome coordination failures}
\label{sec:failure}
Consider the parameter regime where the optimal strategy corresponds to the large-$\eta^s$ solution, in the region where it requires coordination --- that is, $h_{ch} < h < h_M$ and $j>j_B$. If the population fails to coordinate, the actual fraction of buyers at equilibrium will lie on the low-demand curve. May the monopolist drive it to the optimal profit solution, on the high-$\eta$ branch?

Anticipating the risk of coordination failure, the seller may then first post a  price low enough to reach the large-$\eta^d$ customers' equilibrium outside the 
multiple-solutions region (see figures \ref{fig:examplesj35_h-12_h-14_h-15}  for graphical illustrations and Section \ref{sec:Numerical_illustrations} for further discussions), and then smoothly increase the price to reach its optimal value on this branch. If the monopolist had enough econometric measures allowing to precisely estimate the relevant parameters, he should post an introductory price just below $p^*=h+\hat p_L(j)$, so that $\eta$ would match the single Nash equilibrium existing for the demand, which for these parameters lies on the high-$\eta^d$ branch. Clearly, this is possible only if $h>-\hat p_L$, as in figures \ref{fig:examplesj35_h-12_h-14_h-15}(b) and (d), but not in the case of figure \ref{fig:examplesj35_h-12_h-14_h-15}(f) because $h$, the mean IWP, is too negative. Then, the monopolist would be able to drive smoothly the system to the point maximizing his profit, with slightly less 
demand and higher prices. This introductory pricing strategy 
is of the kind discussed by \cite{CabralSalantWoroch99} and could be considered within the two-period pricing studied in \cite{BensaidLesne96}. 

Alternatively, 
the monopolist may begin by posting the price that would maximize his profit if the customers 
did coordinate.
 For $h$ in the range  $[\max\{h_{ch}(j), - \hat p_L(j)\},\; h_M(j)]$, coordination failure may occur, 
in which case the demand will find itself on 
the low-$\eta$ suboptimal branch. The seller can then drive the population to the $L$ boundary by smoothly decreasing the price until finding the point $p^*$. Upon a further small price reduction the fraction of buyers will suddenly jump to the high-$\eta$ equilibrium. From that point on the price may be smoothly increased until reaching its optimal value on the high-$\eta$ branch. 
As shown in Appendix \ref{sec:max_boundaries}, the profit 
is a decreasing function of the price at the boundary $L$ of the low demand branch:
for $h> - \hat p_L(j)$ there is a
maximum of the profit at $\eta_L$, the $L$ boundary of the low-$\eta$ branch --- in particular
for $h> h_-$, this is the unique maximum on this low-$\eta$ branch (see Figure \ref{fig:pi_vs_p}, dot-dashed line ($h=0$), and Figure \ref{fig:examplesj35_h-12_h-14_h-15}, (a) and (b)).  
This is not a stable 
equilibrium since precisely 
if the price decreases below the value corresponding
to this maximum 
the demand jumps to the high-demand manifold because the low-$\eta^d$ equilibrium disappears.
But this remark implies that, for $\max\{h_{ch}(j), -\hat p_L(j)\} < h < h_M$, 
if the population does not coordinate and the demand remains in the low-$\eta^d$ branch,
the monopolist may {\em increase} his profit by decreasing the price, driving the customers 
to the $L$ boundary. In such case (Figure \ref{fig:pi_vs_p}, $h=0$), even without being aware of the existence of the high-$\eta$ branch, starting from the low-$\eta$ suboptimal branch with a relatively high posted price, a t\^atonnement strategy with smooth modifications of the price may allow the monopolist drive the demand onto the high-$\eta$ branch, and then reach the maximal profit.  
However, for $ h_{ch}(j) < h < -\hat p_L(j)$, none of these scenarii 
may be implemented: in that case the $L$ boundary
is in the domain of negative values of $\hat p$  (figures \ref{fig:examplesj35_h-12_h-14_h-15} (e) and (f) ). 
Hence,
the largest value of $\eta$ that the monopoly may reach, even at a vanishing $\hat p$, lies inside the  
multiple demand region. Attracting enough customers to draw the system outside this region would require posting prices below the production cost. Under such conditions, that is if $h< - \hat p_L(j)$, neither the introductory pricing nor the t\^atonnement strategy are implementable:  the risk due to the possibility of coordination failure 
cannot be avoided.

\subsection{Minimax regret strategy}
\label{sec:becker}
Recall that the high-$\eta$ optimal solution is vulnerable not only because
it requires coordination from the customers, but also  
due to its sensibility to small changes in the parameter values: this optimum  
lies at values of $\eta$ very close to the $U$-boundary, that is the point where the branch of high-$\eta$ demand  disappears. 

If the average willingness to pay is negative and the social interactions are strong enough, a suitable risk averse policy 
should 
rely neither on the hypothesis that customers will coordinate, nor on the expectation that if coordination does occur, it will maintain itself. 
A risk-averse monopolist should thus choose a least regret strategy, posting the price corresponding to a sub-optimal profit, either targeting a fraction  $\eta^s_-$ of buyers, i.e. the local maximum on the low-$\eta^d$ branch, or a supply on the high-$\eta$ branch just outside the multi-valued domain ($\eta$ slightly larger than $\eta_L$), whichever gives the highest profit -- or is less risky. 

In the first case, the restaurant is thus designed to have a low capacity. If the system's parameters lie in the region where customers have two equilibria (region with the double arrow in figure \ref{fig:MonopolistPhD_LowEta}) and it happens that it becomes ``in", i.e. the customers coordinate on the high-$\eta^d$ equilibrium, there will be a heavy demand, larger than the capacity, hence a queue for tables. 

In the second case -- admissible only if the $L$-boundary is in the domain of positive values of $\hat p$--, the seller tries to keep the price at the value for which the demand jumped to the high-$\eta$ branch, just outside the domain with multi-valued demand. If a change occurs in the population characteristics, 
so that the current posted price finds itself back into this domain, a failure in coordination can be
cured by a small price decrease, finding again the $L$-boundary.

\section{Conclusion}
\label{sec:conclusion}
We analyzed the monopolist's profit optimization problem when there exist social influences among customers. We considered the generic case of finite variance distributions of the idiosyncratic willingness to pay (IWP) of the customers population. The model suggests possible explanations to the fact that prices of goods with bandwagon effects do not necessarily increase despite their success.

There are two critical (normalized) interaction strengths $j_A$ and $j_B$ whose precise values depend on the details of the IWP distribution. Depending on the strength $j$ of the (normalized) social interactions the monopoly may be faced with three different 
situations:  
for $j<j_A$, 
the profit optimization gives a single price for each value of the (normalized) average willingness to pay of the population, $h$, and for this posted price the demand is single valued. For $j > j_A$, there is a range of $h$ values where the monopolist's profit presents two maxima. There is a first order transition where the optimal strategy shift from selling  to few customers at a high price, to selling to a large fraction of customers at a low price. 
If $j_A < j < j_B$, the customers system equilibrium is a single valued function of the price. Hence the monopolist can drive the market through the posted prices, and thus earn the maximum profit. The situation is very different for $j>j_B$ 
when the average willingness to pay is small enough -- it may be smaller than the monopolist's cost (i.e. $h<0$). 
In this region,
the customers meet a coordination problem and the monopolist cannot drive the market though prices alone. In addition, the price that is expected to give the largest profit is generically just below the critical value where the high demand equilibrium disappears, implying the risk of a collapse of the demand from  unexpected changes in the customers characteristics. Social interactions thus introduce uncertainty in the outcome of the monopolist strategy. 
In our discussion we presented different pricing strategies targeting the optimal or a suboptimal profit under these conditions.

The introductory pricing strategy is 
discussed in section \ref{sec:failure}. 
We suggest another possible explanation for Becker's restaurant pricing problem. If the restaurant determines its optimal price, and it happens to lie on the high demand 
branch, it will propose the corresponding number of tables. 
If this equilibrium lies within  the customers' coordination region, posting as introductory price 
a price just below the domain of multiple solutions will attract more customers than seats, with the corresponding queue at the entrance. Lasting of this situation reveals a precautionary risk dominant pricing strategy. 

Many extensions of the analysis could be considered. First, the discussion on how the seller could drive the customers to a domain with higher profit might be recast within the framework of control theory, with a cost function taking into account the cumulative gain with a discount rate. Such an approach has already been developed for problems where multiple equilibria exist, see \cite{LakeGame}.  Second, in our analysis we have assumed non strategic behaviour from the customers with respect to the seller. It will be interested to consider at least two periods, where some customers may anticipate a drop of prices at the second period, and thus may wait for buying, as it occurs for high-tech goods. The seller's strategy would then have to be adapted accordingly. In \cite{BensaidLesne96} a similar problem is studied, but in a market situtation where, on the countrary, prices are expected to increase with time. Finally, the case of an oligopolistic market should be studied.

\subsubsection*{Acknowledgements} 
We are grateful to David Martimort for helpful suggestions. We thank Annick Vignes for useful comments and for a critical reading of the manuscript. We gratefully acknowledge the anonymous referees for their comments. This work was part of the 
project ELICCIR supported by
the joint programme "Complex Systems in Human and 
Social Sciences" of the French Ministry of Research
and of the CNRS, and of the project DyXi supported by the Programme SYSCOMM of the French National Research Agency, the ANR (grant ANR-08-SYSC-008). 
M.B.G., J.-P.N. and D.P. are CNRS members.


\begin{thebibliography}{10}

\bibitem{Arthur94}
W.~B. Arthur.
\newblock El Farol.
\newblock {\em Amer. Econ. Review}, 84:406, 1994.

\bibitem{Arthur99}
W.~B. Arthur.
\newblock Complexity and the economy.
\newblock {\em Science}, 284:107--109, 1999.

\bibitem{Art73}
R.~Artle and C.~Averous.
\newblock The telephone system as a public good: Static and dynamic aspects.
\newblock {\em The Bell Journal of Economics and Management Science},
  4(1):89--100, 1973.

\bibitem{BensaidLesne96}
B.~Bansaid and J.-Ph. Lesne.
\newblock Dynamic monopoly pricing with network externalities.
\newblock {\em International Journal of Industrial Organization},
  14(6):837--855, 1996.

\bibitem{Becker91}
G.~S. Becker.
\newblock A note on restaurant pricing and other examples of social influences
  on price.
\newblock {\em The Journal of Political Economy}, 99:1109--1116, 1991.

\bibitem{BeckerMurphy00}
G.~S. Becker and K.M Murphy.
\newblock {\em Social Economics. Market Behavior in a Social Environment}.
\newblock Cambrige Ma. The Belknap Press \& Harvard University Press, 2000.

\bibitem{Ben96}
R.~Benabou.
\newblock Equity and efficiency in human capital investment: The local
  connection.
\newblock {\em Review of Economic Studies}, 63:237--264, 1996.

\bibitem{BorghesiBouchaud}
C.~Borghesi and J.-P. Bouchaud.
\newblock Of songs and men: a model for multiple choice with herding.
\newblock {\em Quality and Quantity}, 41(4):557--568, 2007.

\bibitem{BouchaudReview2012}
J.-Ph. Bouchaud.
\newblock Crises and collective socio-economic phenomena: cartoon models and
  challenges.
\newblock {\em arXiv:1209.0453v1}, 2012.

\bibitem{BrockDurlauf01a}
W.~A. Brock and S.~N. Durlauf.
\newblock Discrete choice with social interactions.
\newblock {\em Review of Economic Studies}, 68:235--260, 2001.

\bibitem{CabralSalantWoroch99}
L.~M.~B. Cabral, D.~J. Salant, and G.~A. Woroch.
\newblock Monopoly pricing with network externalities.
\newblock {\em International Journal of Industrial Organization},
  17(2):199--214, 1999.

\bibitem{ConleyTopa02}
T.~Conley and G.~Topa.
\newblock Socio-economic distance and spatial patterns in unemployment.
\newblock {\em Journal of Applied Econometrics}, 17(4):303--327, 2002.

\bibitem{Crane91}
J.~Crane.
\newblock The epidemic theory of ghettos and neighborhood effects of dropping
  out and teenage childbearing.
\newblock {\em American Journal of Sociology}, 96:1226--1259, 1991.

\bibitem{Cur87}
N.~Curien and M.~Gensollen.
\newblock Les th\'eories de la demande de raccordement t\'el\'ephonique.
\newblock {\em Revue Economique}, Vol. 2 mars 1987:203--255, 1987.

\bibitem{LakeGame}
W.~D. Dechert and W.~Brock.
\newblock The lake game,.
\newblock In C.~Perrings K.-G.~M?aler and D.~Starrett, editors, {\em Economic
  and Ecological Modelling}. The Beijer Institute, The Royal Academy of
  Sciences, Stockholm, 2000.

\bibitem{Durlauf96}
S.~N. Durlauf.
\newblock A theory of persistent income inequality.
\newblock {\em Journal of Economic Growth}, 1:349--366, 1996.

\bibitem{Durlauf97}
S.~N. Durlauf.
\newblock Statistical mechanics approaches to socioeconomic behavior.
\newblock In B.~Arthur, S.~N. Durlauf, and D.~Lane, editors, {\em The Economy
  as an Evolving Complex System II}, pages 81--104. Santa Fe Institute Studies
  in the Sciences of Complexity, Volume XVII, Addison-Wesley Pub. Co, 1997.

\bibitem{Durlauf06}
S.~N. Durlauf.
\newblock Groups, social influences, and inequality: A membership theory
  perspective on poverty traps.
\newblock In S.~Bowles, S.~N. Durlauf, and K.~Hoff, editors, {\em Poverty
  Traps}. Princeton: Princeton University Press, 2006.

\bibitem{Follmer74}
H.~F{\"{o}}llmer.
\newblock Random economies with many interacting agents.
\newblock {\em Journal of Mathematical Economics}, 1(1):51--62, 1974.

\bibitem{GaGeSh}
S.~Galam, Y.~Gefen, and Y.~Shapir.
\newblock Sociophysics: A mean behavior model for the process of strike.
\newblock {\em Mathematical Journal of Sociology}, {\bf 9}:1--13, (1982).

\bibitem{GlaeserSacerdoteScheinkman96}
E.~L. Glaeser, B.~Sacerdote, and J.~A. Scheinkman.
\newblock Crime and social interactions.
\newblock {\em Quarterly Journal of Economics}, CXI:507--548, 1996.

\bibitem{GoNaPhVa05}
M.~B. Gordon, J.-P. Nadal, D.~Phan, and J.~Vannimenus.
\newblock Seller's dilemma due to social interactions between customers.
\newblock {\em Physica A}, 356 Issues 2-4:628--640, 2005.

\bibitem{GoNaPhSeM3as}
Mirta~B. Gordon, Jean-Pierre Nadal, Denis Phan, and Viktorija Semeshenko.
\newblock Discrete choices under social influence: Generic properties.
\newblock {\em Mathematical Models and Methods in Applied Sciences (M3AS)},
  19(Supplementary Issue 1):1441--1481, 2009.

\bibitem{Granovetter78}
M.~Granovetter.
\newblock Threshold models of collective behavior.
\newblock {\em American Journal of Sociology}, 83(6):1360--1380, 1978.

\bibitem{GranovetterSoong86}
M.~Granovetter and R.~Soong.
\newblock Threshold models of interpersonal effects in consumer demand.
\newblock {\em Journal of Economic Behavior and Organization}, 7:83--99, 1986.

\bibitem{IoannidesZabel03}
Y.M. Ioannides and J.~Zabel.
\newblock Neighborhood effects and housing demand.
\newblock {\em Journal of Applied Econometrics}, 18:563--584, 2003.

\bibitem{KatzShapiro94}
M.~L. Katz and C.~Shapiro.
\newblock Systems competition and network effects.
\newblock {\em The Journal of Economic Perspectives}, 8(2):93--115, 1994.

\bibitem{Krauth06a}
B.~Krauth.
\newblock Social interactions in small groups.
\newblock {\em Canadian Journal of Economics}, 39(2):414--433, 2006.

\bibitem{Krauth06b}
B.~V. Krauth.
\newblock Simulation-based estimation of peer effects.
\newblock {\em Journal of Econometrics}, 133(1):243--271, July 2006.

\bibitem{Leibenstein50}
H.~Leibenstein.
\newblock Bandwagon, snob, and veblen effects in the theory of consumers'
  demand.
\newblock {\em Quarterly Journal of Economics}, 64(2):183--207, 1950.

\bibitem{Manski00}
C.~F. Manski.
\newblock Economic analysis of social interactions.
\newblock {\em The Journal of Economic Perspectives}, 14(3):115--136, 2000.

\bibitem{MichardBouchaud}
Q.~Michard and J.-P. Bouchaud.
\newblock Theory of collective opinion shifts: from smooth trends to abrupt
  swings.
\newblock {\em The European Physical Journal B - Condensed Matter and Complex
  Systems}, 47:151--159, 2005.

\bibitem{NaPhGoVa05}
J.-P. Nadal, D.~Phan, M.~B. Gordon, and J.~Vannimenus.
\newblock Multiple equilibria in a monopoly market with heterogeneous agents
  and externalities.
\newblock {\em Quantitative Finance}, 5(6):557--568, 2005.

\bibitem{Orlean95}
A.~Orl{\'e}an.
\newblock Bayesian interactions and collective dynamics of opinion: Herd
  behaviour and mimetic contagion.
\newblock {\em Journal of Economic Behavior and Organization}, 28:257--274,
  1995.

\bibitem{Ostrom00}
E.~Ostrom.
\newblock Collective action and the behaviour of social norms.
\newblock {\em Journal of Economic Perspectives}, 14:137--158, 2000.

\bibitem{PhaGorNad04}
D.~Phan, M.~B. Gordon, and J.-P. Nadal.
\newblock Social interactions in economic theory: an insight from statistical
  mechanics.
\newblock In Bourgine P. and Nadal J-P., editors, {\em Cognitive Economics},
  pages 335--358. Springer, 2004.

\bibitem{Roh74}
J.~Rohlfs.
\newblock Theory of interdependent demand for a communications service.
\newblock {\em Bell Journal of Economics and Management Science}, Vol. 5 No. 1
  Spring:16--37, 1974.

\bibitem{Rohlfs01}
J.~Rohlfs.
\newblock {\em Bandwagon effects in High-Technology Industries}.
\newblock Cambridge, Ma., MIT Press, 2001.

\bibitem{Schelling71}
T.~C. Schelling.
\newblock Dynamic models of segregation.
\newblock {\em Journal of Mathematical Sociology}, 1:143--186, 1971.

\bibitem{Schelling73}
T.~C. Schelling.
\newblock Hockey helmets, concealed weapons, and daylight saving: A study of
  binary choices with externalities.
\newblock {\em The Journal of Conflict Resolution}, 17(3):381--428, September
  1973.

\bibitem{Schelling}
T.~S. Schelling.
\newblock {\em Micromotives and Macrobehavior}.
\newblock W.W. Norton and Co, N.LY., 1978.

\bibitem{SeGoNa08}
Viktoriya Semeshenko, Mirta~B. Gordon, and Jean-Pierre Nadal.
\newblock Collective states in social systems with interacting learning agents.
\newblock {\em Physica A}, 387:4903--4916, 2008.

\bibitem{Sethna}
J.~P. Sethna.
\newblock {\em Statistical Mechanics: Entropy, Order Parameters and
  Complexity}.
\newblock Oxford Univ. Press, 2006.

\bibitem{ShapVar99}
C.~Shapiro and H.~Varian.
\newblock {\em Information Rules: A Strategic Guide to the Network Economy}.
\newblock Harvard Business School Press, Boston, Massachusetts, 1999.

\bibitem{SoeteventKooreman06}
A.R. Soetevent and P.~Kooreman.
\newblock A discrete choice model with social interactions; with an application
  to high school teen behavior.
\newblock {\em Journal of Applied Econometrics}, 22(3):599--624, April/May
  2007.

\bibitem{Topa01}
G.~Topa.
\newblock Social interactions, local spillovers and unemployment.
\newblock {\em Review of Economic Studies}, 68(2):261--295, 2001.

\bibitem{vRab74}
B.~Von~Rabenau and K.~Stahl.
\newblock Dynamic aspects of public goods: a further analysis of the telephone
  system.
\newblock {\em Bell Journal of Economics and Management Science}, Vol. 5
  n2:651--669, 1974.

\bibitem{WeiSta}
G.~Weisbuch and D.~Stauffer.
\newblock Adjustment and social choice.
\newblock {\em Physica A}, 323:651--662, 2003.

\end{thebibliography}

\clearpage
\pagebreak

\renewcommand{\theequation}{A-\arabic{equation}} 
\setcounter{equation}{0}  
\renewcommand{\thesection}{A} 
\setcounter{section}{0}  
\renewcommand{\thefigure}{A-\arabic{figure}} 
\setcounter{figure}{0}  

\section{Appendices}
\label{sec:appendices}

\subsection{Demand equilibria from the function ${\cal D}$}
\label{sec_app:graphic}

\begin{figure}[ht]
\centering
\includegraphics[width=0.60\textwidth]{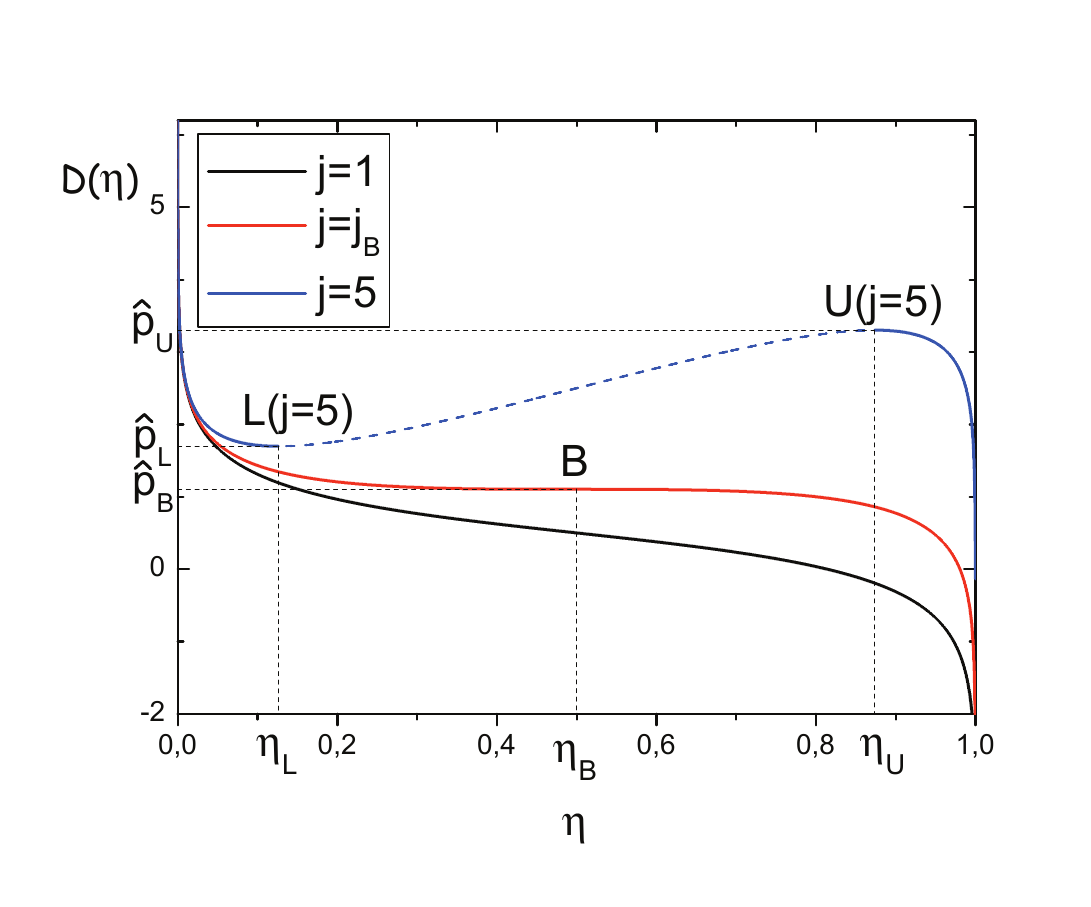}
\caption[]{\em For the case of a logistic distribution of the IWP, graph of the function ${\cal D}$ for $j=1<j_B$ (classical behaviour at low social influence or high mean willingness to pay), $j=j_B \approx 2.21$  (curve passing through the point B), and $j=5 > j_B $ (multiply valued function at large social influence; dashed: unstable equilibria).}
\label{fig_app:inversedemand_D}
\end{figure} 

The demand equilibria $\eta^d(\hat p,j)$ are the solutions to (\ref{eq:p-h}). 
Graphically, at a given value of $j$, they correspond to the intersection(s) of the function $y={\cal D}(j;\eta)$ with the horizontal line $y=\hat p$. Plots of ${\cal D}(j;\eta)$ against $\eta$ for different values of $j$ are presented on Figure \ref{fig_app:inversedemand_D}.  According to Equ. (\ref{eq:p^d}), these curves, shifted vertically by the value $h$ (the average IWP of the population) give the  demand curves like those represented on Fig.~\ref{fig:inversedemand_D}.

\subsection{Profit maximization}
\label{sec:price_opt}

\subsubsection{An effective supply function}
\label{sec_app:effectivesupply}

There are different equivalent ways of performing the profit maximization when social interactions are present. One is to consider that $\eta$, and consequently the profit, are functions of $p$ (as in \cite{NaPhGoVa05}): $\eta 
=\eta^d(p)$ , and $\pi(p)= p \; \eta^d(p)$ ; another one is to consider that $p$
 and the profit are functions of $\eta$, $p=p^d(\eta)$, and $\pi(\eta)= 
p^d(\eta) \; \eta$. Here we follow a reasoning that leads (obviously) to the same results, but deals symmetrically with the variables $p$ and $\eta$. The interest of this approach is that it puts forward an analogy between the demand and the supply equations. 

In order to maximize the monopolist's profit, let us define
\begin{equation}
\Psi(\eta,p) \equiv p^d(\eta) - p.
\label{eq:implicit} 
\end{equation}
The equation  $\Psi(\eta,p) = 0$ defines a curve $\Psi$ in the plane $\{\eta,p\}$ along which $\pi$ has to be maximized. Be 
\begin{equation}
{\bf v}(\eta,p) \equiv (v_\eta,v_p)=(\partial \Psi/ \partial p,- \partial \Psi/ \partial \eta) 
\label{eq:tangent}
\end{equation}
a vector tangent to the curve $\Psi =0$ at the point $(\eta,p)$. 

The maximization of (\ref{eq:profit2}) along $\Psi$ imposes that the directional derivative of $\pi$ vanishes,
\begin{equation}
({\bf v} \cdot {\nabla}) \; \pi \equiv v_{\eta} \frac{\partial \pi}{\partial \eta} + v_p \frac{\partial \pi}{\partial p} \; = \; 0,
\label{eq:directional_deriv}
\end{equation}
to guarantee that the profit is an extremum. If the maximum is reached inside the support of $\Gamma(\eta)$, it must also satisfy the second order condition
\begin{equation}
({\bf v} \cdot {\nabla})\; (v_{\eta} \frac{\partial \pi}{\partial \eta} + v_p \frac{\partial \pi}{\partial p})\leq 0.
\label{eq:directional_max}
\end{equation} 
For finite range pdfs (compact support), one has to check whether the maximum maximorum lies 
on one of the boundaries of the support. Remark: had we chosen to consider $\eta$ as a function of $p$, 
that is $\pi(p) = p \eta^d(p)$ as done in~\cite{NaPhGoVa05}, this stability condition would read $$ \frac{d^2 \pi}{d p^2} \leq 0. $$ 

Introducing the components of $\bf v$ given by equation (\ref{eq:tangent})
\begin{eqnarray}
v_\eta &=& \frac{\partial \Psi}{\partial p} = -1,   \label{eq:comp_v_eta} \\
v_p &=& -\frac{\partial \Psi}{\partial \eta}=-\; \frac{dp^d(\eta)}{d\eta}, \label{eq:comp_v_p}
\end{eqnarray}
into the first order condition (\ref{eq:directional_deriv}) gives 
\begin{equation}
\label{eq:p_monop}
p^d(\eta)=p^s(\eta)
\end{equation}
where $p^d(\eta)$ is the inverse 
demand given by equation (\ref{eq:p^d}), and
\begin{equation}
\label{eq:ps}
p^s(\eta) \equiv -\eta \frac{dp^d(\eta)}{d\eta},
\end{equation}
which, making use of the expression (\ref{eq:p^d}) of $p^d$, is also given by
\begin{equation}
p^s(\eta) = -\eta {\cal D}'(j;\eta) = \eta [\Gamma'(\eta)-j] .
\label{eq:ps2}
\end{equation}

Using (\ref{eq:comp_v_eta}), (\ref{eq:comp_v_p}) and (\ref{eq:ps}), the second order condition (\ref{eq:directional_max}) reads
\begin{equation}
[- \frac{\partial}{\partial \eta} - \frac{d p^d(\eta)}{d\eta} \frac{\partial}{\partial p}]
\;[-p + p^s(\eta)] \; \leq \;0,
\label{eq:2ndordercond_appendix}
\end{equation}
and this can be rewritten as
\begin{equation}
\frac{d }{d\eta} [p^d(\eta) - p^s(\eta)] \leq 0.
\label{eq:2ndordercond}
\end{equation}

We may consider $p^s(\eta)$ given by (\ref{eq:ps}) as an effective inverse 
supply function, although it is clearly not a true one, since it is defined by the monopolist's optimization program, itself based on the knowledge of the demand function. 
Nevertheless, it has all the properties of an inverse 
supply function, and the market equilibrium can be understood from the equality (\ref{eq:p_monop}) between demand and (effective) supply.
Note that from (\ref{eq:ps}) the positivity of the supply price $p^s$ is equivalent to having the inverse 
demand decreasing with $\eta$ (or equivalently, from (\ref{eq:ps2}), to have the demand at a stable equilibrium, that is ${\cal D}'(j; \eta) \le 0$.

\subsubsection{Behaviour of $\widetilde{\Gamma}$}
\label{sec:app_Gamma_hat}
In order to have a full analogy between the Supply and Demand problems, on needs $\widetilde{\Gamma}(\eta)$, defined by Eq (\ref{eq:hatgamma}), to be monotonically increasing from $-\infty$ to $+\infty$ when $\eta$ goes from $0$ to $1$, and  to have a single inflexion point. 
One  can easily check that, for `usual' pdfs such as the logistic or the Gaussian,
$\widetilde{\Gamma}$ is a monotonic, strictly increasing function of $\eta$ for all $0 < \eta < 1$. 
For an arbitrary (smooth enough) pdf, one can show that 
 $\widetilde{\Gamma}$ is a monotonically increasing function for $\eta > \eta_B$, as well as for $\eta$ small enough whenever the pdf has a finite variance. 
 It remains the possibility to have a non monotonic behaviour of $\widetilde{\Gamma}$ in some small intermediate range, $\eta$ not too small and not too close to $\eta_B$.

Let us now express the derivative of $\widetilde{\Gamma}$ in term of the pdf $f(x)$ and its cumulative  $G(x) \equiv 1-F(x)$. One has $\eta=G(x)$. From the definition of
$\widetilde{\Gamma}$, one has
$$
\frac{ d \widetilde{\Gamma}}{d \eta } = -2 \frac{ dx }{d\eta }\; - G(x) \frac{ d^2x }{d\eta^2 }.
$$
Now $\frac{ dx }{d\eta }=1/G'(x)$, and thus $\frac{ d^2x }{d\eta^2 } = -\frac{G''}{G'^3}$.
It follows that one can write 

$$\frac{ d \widetilde{\Gamma}}{d \eta } \; = \; -\frac{G^3}{G'^3} \; \left[ -\frac{G''}{G^2} + 2 \frac{G'^2}{G^3} \right]
\; = \;-[\frac{G}{G'}]^3 \; \left[ \frac{ d^2}{dx^2 } \; \frac{ 1}{G(x) } \right]$$

which means
\begin{equation}
\frac{ d \widetilde{\Gamma}}{d \eta } = [\frac{1-F(x)}{f(x)}]^3 \; \left[ \frac{ d^2}{dx^2 } \; \frac{ 1}{1-F(x) } \right]
\label{eq:gammahat}
\end{equation}
The right hand side is positive if and only if
\begin{equation}
\frac{ d^2 }{dx^2 }\; \frac{ 1 }{1-F(x) } \; > \; 0.
\label{eq_app:cond_on_f_b}
\end{equation}
Hence $\widetilde{\Gamma}$ is a monotonic, strictly increasing function of $\eta$ for all $0 < \eta < 1$ iff the above property 
is true for every $x$ belonging to the support of $f$.

This condition (\ref{eq_app:cond_on_f_b}) is not a very stringent one. After some algebra it may be shown that it is equivalent to impose that, {\it in the absence of externalities}  ($j=0$), the demand satisfies  
\begin{equation}
\frac{ d^2 }{ dp^2 }\; \frac{ 1 }{\eta^d(p)}\; > \; 0.
\end{equation}
This is a weaker condition than a convexity condition,  $d^2 \log \eta^d(p) /dp^2<0$, frequently assumed in economics.

\subsection{Numerical illustrations}
\label{sec:Numerical_illustrations}
\begin{figure}[!tp]
\centering
\begin{tabular*}{\textwidth}{cc}   
\includegraphics[width=0.47\textwidth]{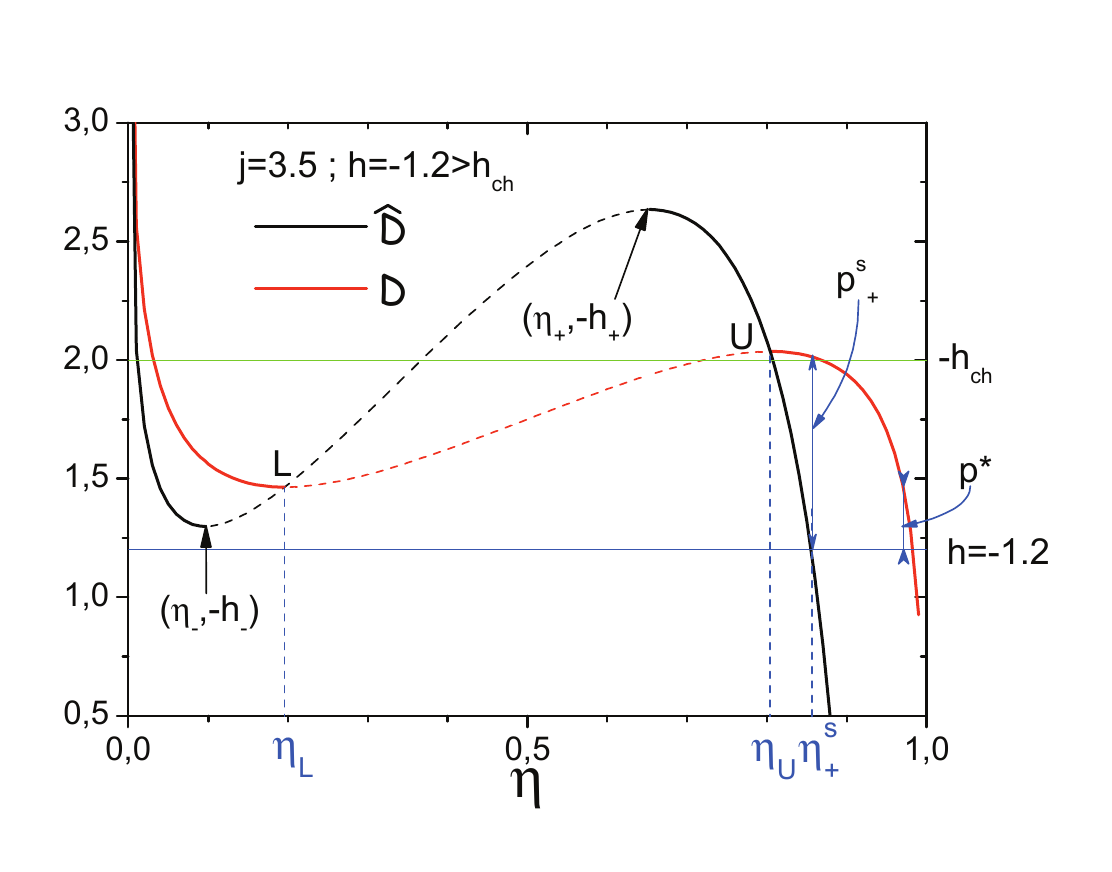} & \includegraphics[width=0.47\textwidth]{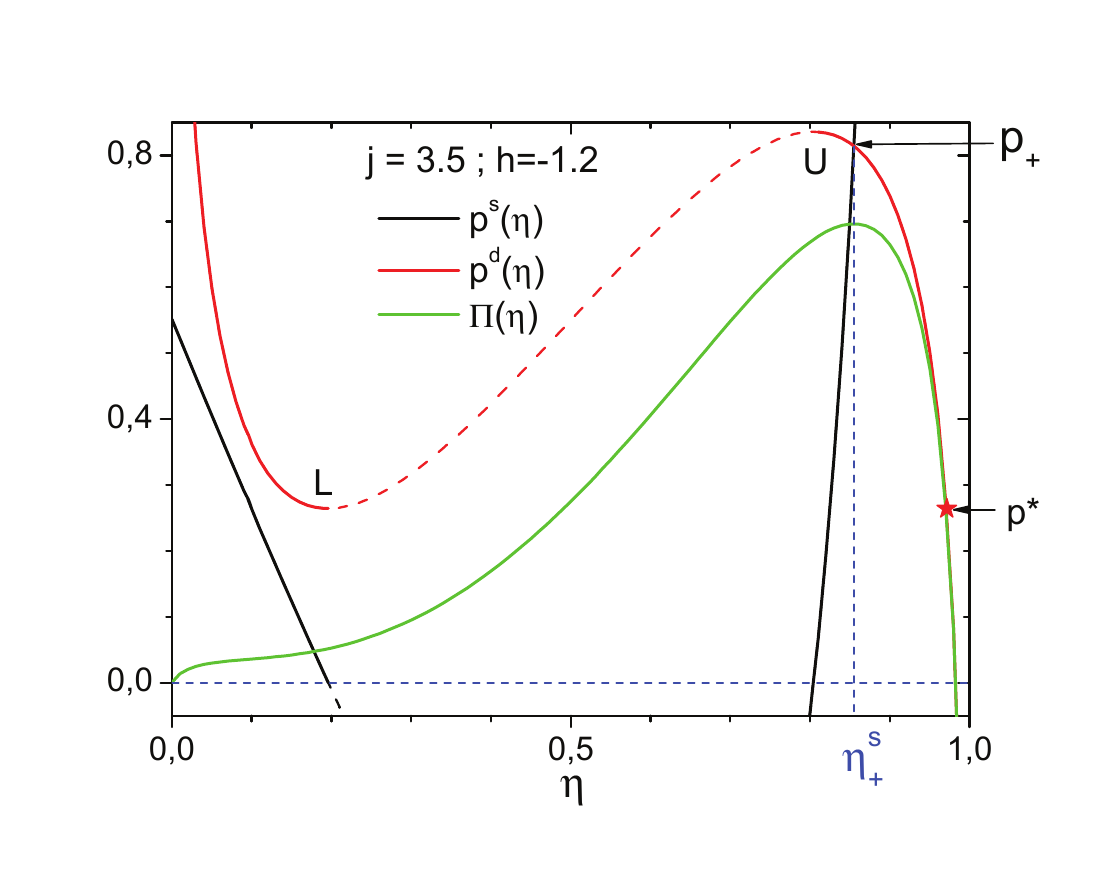} \\
(a) \& (b)\\
\includegraphics[width=0.47\textwidth]{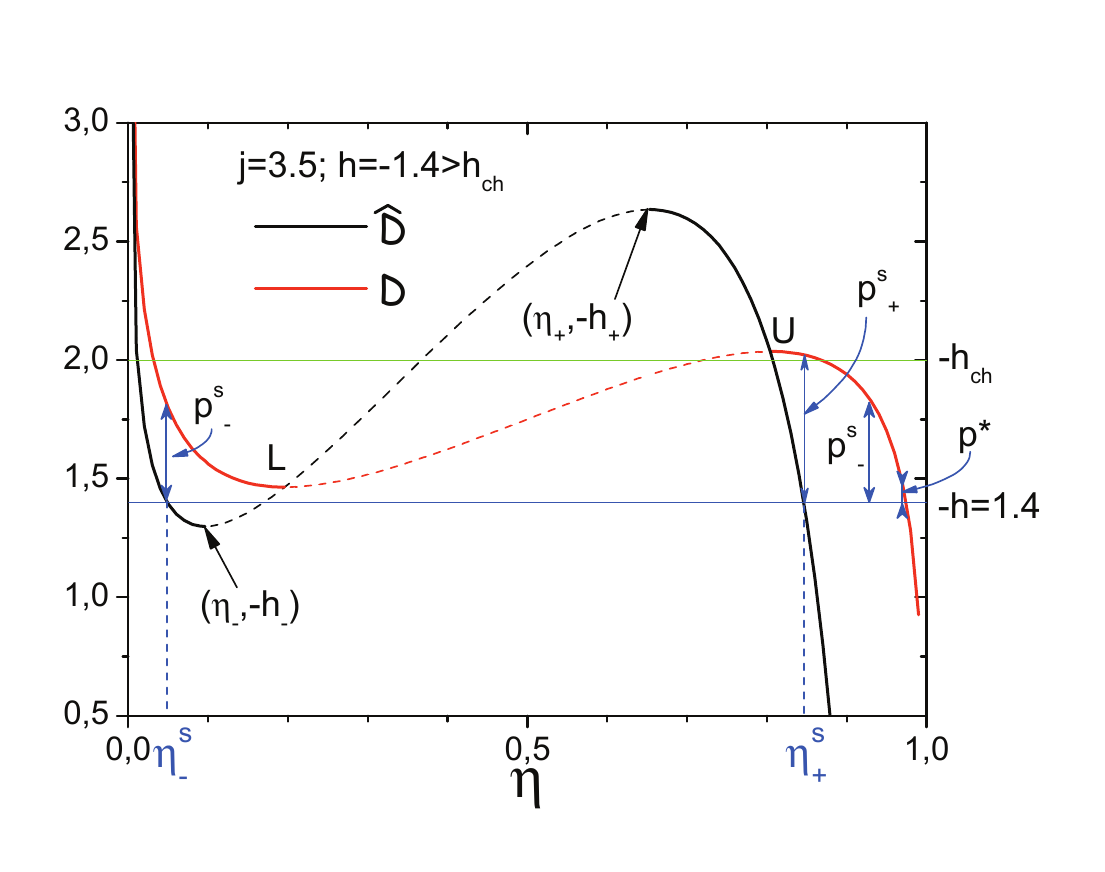} & \includegraphics[width=0.47\textwidth]{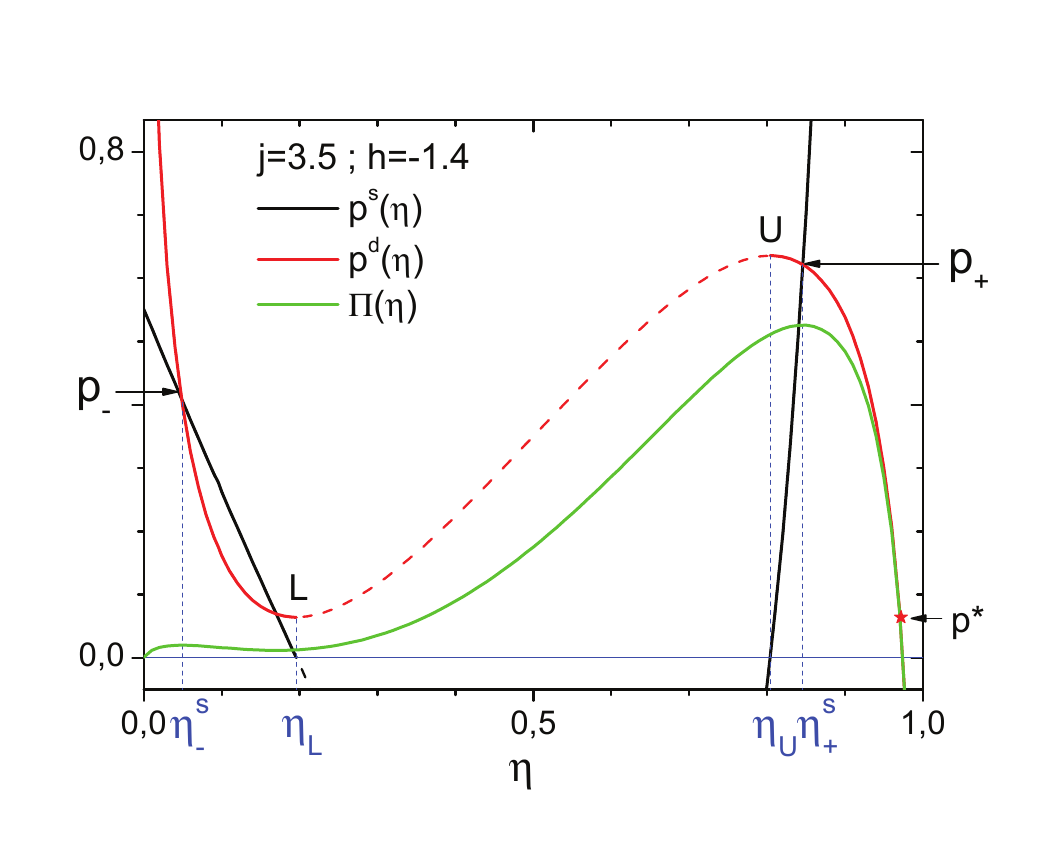} \\
(c) \& (d)\\
\includegraphics[width=0.47\textwidth]{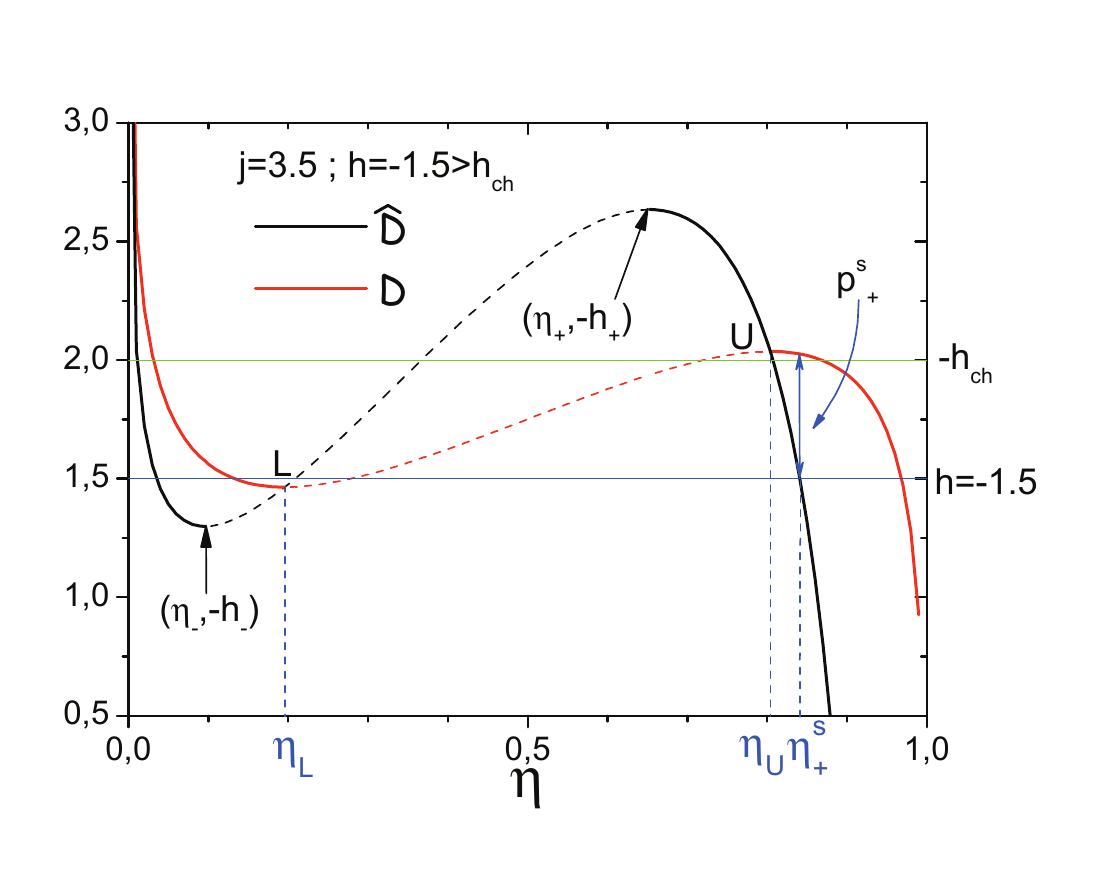} & \includegraphics[width=0.47\textwidth]{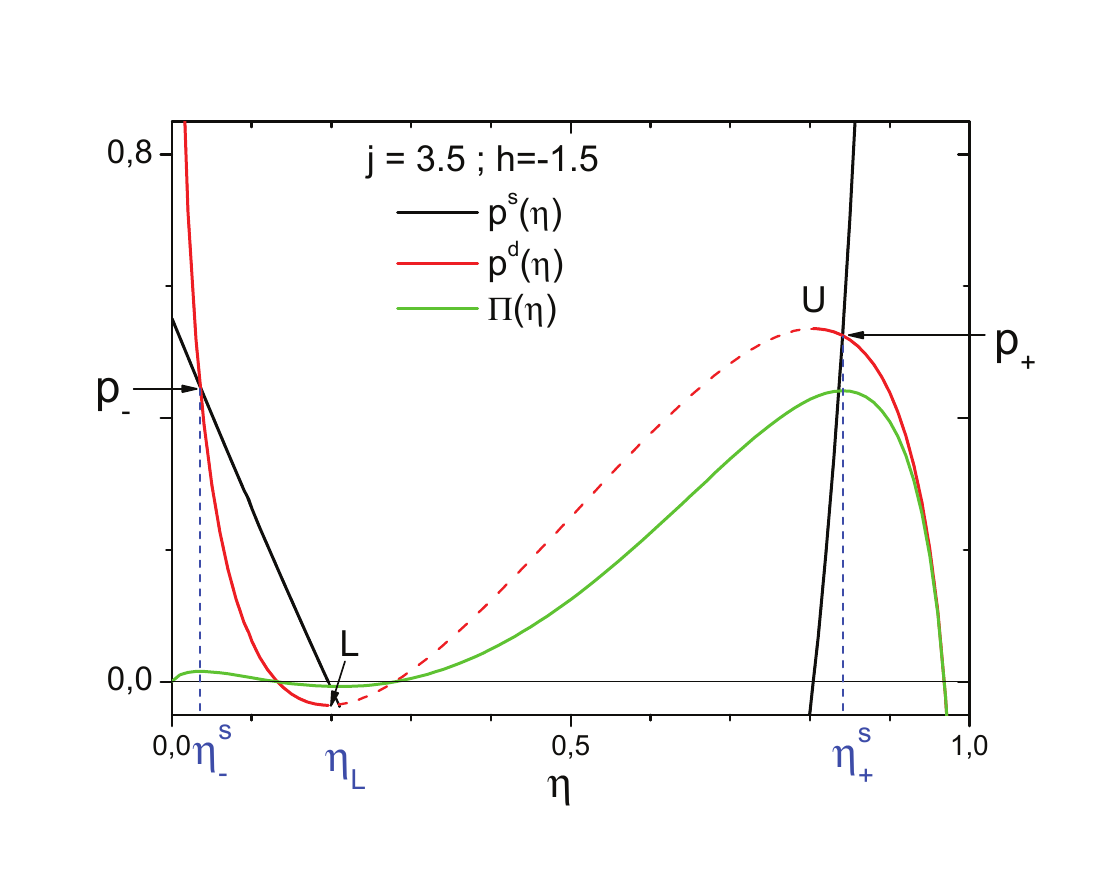} \\
(e) \& (f)
\end{tabular*}
\caption[]{{\em 
Figures (a), (c) and (e): functions ${\cal D}(j; \eta)$ and
$\widetilde{\cal D}(j; \eta)$. The construction that determines the 
profit extrema is illustrated. 
Figures (b), (d) and (f): price and profit vs. $\eta$ showing the extrema of $\pi$. When $h>h_{ch}\approx -2$: the large-$\eta$ strategy is the optimum. The dotted lines correspond to the non-economic demand between $\eta_L$ and $\eta_U$, where ${\cal D}$ has positive slope. }}
\label{fig:examplesj35_h-12_h-14_h-15}
\end{figure}

Figures \ref{fig:examplesj35_h-12_h-14_h-15} present examples of possible situations met by the monopoly. They correspond to systems with parameters $j=3.5>j_B$ (for the logistic, $j_B \approx 2.21$)  
and the three different values of $h$, indicated on figure \protect{\ref{fig:MonopolistCompletePhD}}, that fall within the region of uncertain outcome for the $\eta^s_+$ strategy. 
The curves $\cal D$ and $\widetilde{\cal D}$ on Figures (a), (c) and (e) are typical of any smooth pdf in the region with multiple solutions of the profit optimization. They
show the constructions allowing to determine the optimal prices. The lines $y=-h_{ch}$ are represented: for the values of $h$ considered the optimal strategy corresponds to the high-$\eta^s$ branch. Optimal prices are given by the difference ${\cal D - \widetilde{\cal D}}$ at the value of $\eta$ where the line $y=-h$ intersects the curve $\widetilde{\cal D}$. The introductory price $p^*$ is indicated whenever it is viable (i.e. positive). In particular, when $h=-1.5$ introductory prices allowing to get rid of the coordination problem do not exist. Figures (b), (d) and (f) present the corresponding demand and supply prices, $p^d=h+\cal D$ and $p^s$ (the optimal price corresponds to $p^s=p^d$), and the profits $\pi=\eta p^s$. In (a,b) the only optimal strategy is $\eta^s_+$. In (c,d) both strategies $\eta^s_-$ and $\eta^s_+$ fall inside the 
multi-valued demand region. The profit of targeting $\eta^s_-$, which does not need coordination, is smaller than the profit corresponding to the introductory price $p^*$, and the latter may allow to drive the system to the $\eta^s_+$ equilibrium of optimal profit. If $h=-1.5$, figures (e,f), both strategies fall in the coordination region, and there is no possibility to get rid of the problem through introductory prices.

\subsection{Monopolist's Phase Diagram: Details}
\label{sec:app_phasediag}

\subsubsection{Domain of multiple pricing strategies} 
\label{sec:app_dilemma}
Like for the demand, there is for the supply a bifurcation at critical value of $j$ defined by 
\begin{equation}
j_A=\widetilde{\Gamma}'(\eta_A)/2, \;\;\;\; {\rm with} \;\;\;\;\eta_A \equiv \arg \min_{\eta} \widetilde{\Gamma}'(\eta),
\label{eq_app:j_A}
\end{equation} 
beyond which there are multiple solutions for the supply. 
One can show that $\eta_A  \leq  \eta_B$ and  $j_A  \leq  j_B$. 

If $j<j_A$ both curves $\cal D$ and $\widetilde {\cal D}$ are monotonically 
decreasing functions of $\eta$. Equation (\ref{eq:h_monop1}) has a single 
solution $\eta^s$ for each couple $j$, $h$. Given any $j (<j_A)$, if $h$ 
increases continuously from a very small value (highly negative) to a large 
positive one, the optimal price --which by (\ref{eq:p^s}) is the difference between $\cal D$ and $\widetilde {\cal D}$-- decreases and the fraction of buyers 
increases, both monotonically. 

For $j \ge j_A$, there is a finite range of values of $h$ where $\widetilde{\cal D}(j; \eta)$ presents two extrema: a minimum at $\eta_-(j) < \eta_A$ and a maximum at $\eta_+(j) > \eta_A$. They satisfy $\widetilde{\cal D}'(j; \eta_{\pm}) = 0$. Correspondingly, the profit presents two relative maxima at the intersections of $y=-h$ with the branches of $\widetilde{\cal D}(j; \eta)$ that have negative slope. Notice that the additional intersection at an intermediate value of $\eta$ with the branch having $\widetilde{\cal D}'(j; \eta) > 0$ corresponds to a {\em minimum} of the profit. The profit's absolute maximum has to be determined numerically, through comparison of the profit relative maxima. 

As an example, plots of $\widetilde{\cal D}(j; \eta)$ and ${\cal D}(j; \eta)$ for a logistic pdf and a value $j > j_B$ are presented on Figure \ref{fig:examplesj35_h-12_h-14_h-15}. We have explicitly indicated the values of ${\cal D}(j; \eta)$ corresponding to the unstable equilibria $\eta \in [\eta_L(j),\eta_U(j)]$. The price construction for a particular value of $h$ is exhibited. 

The values of $\eta_+(j)$ and $\eta_-(j)$ are determined following the same 
steps as for the customer's model, and have the same form as equations 
(\ref{eq:dD=0})
which define $\eta_L(j)$ and $\eta_U(j)$,
but with 
$\widetilde{\Gamma}(\eta)$ and $\hat \jmath$ instead of $\Gamma(\eta)$ and $j$ 
respectively. Introducing the values $\eta_+(j)$ and $\eta_-(j)$ into (\ref
{eq:h_monop1}) we obtain:
\begin{equation}
\label{eq:monop_parameterized}
h_{\pm}(j) = -\widetilde{\cal D}(j; \eta_{\pm}(j)). 
\end{equation}
In the plane $\{j, h\}$ the lines $h=h_+(j)$ and  $h=h_-(j)$ represented on 
figure \ref{fig:MonopolistCompletePhD} are the boundaries of a region where 
the profit has multiple (sub)optima. These boundaries merge at the point $A$ 
that satisfies simultaneously $\widetilde{\cal D}'(j; \eta)=0$, and 
$\widetilde{\cal D}''(j; \eta)= 0$, that is
\begin{equation}
A \equiv \{ j_A , \; h_A \equiv \widetilde{\Gamma}(
\eta_A)-\eta_A \widetilde{\Gamma}'(\eta_A) \}.
\label{eq:A}
\end{equation} 

It may be easily checked that $d[h_-(j)-h_+(j)]/dj=2[\eta_+(j)-\eta_-(j)]$, 
meaning that, like in the customers problem, the width of the region with 
multiple extrema increases with $j$ because $\eta_- < \eta_+$. 

\subsubsection{Null price boundary}
When $j>j_B$ both $\cal D$ and $\widetilde {\cal D}$ present positive slopes, 
but for different ranges of $\eta$. As already stated, both curves cross each 
other at $\eta_L$ and $\eta_U$, with $\widetilde {\cal D}'(\eta_L)>0$ and 
$\widetilde {\cal D}'(\eta_U)<0$. Thus, at $\eta_L$ the profit is a minimum. 
Since $\eta_-<\eta_L$, for any $\eta^s<\eta_-$, $\widetilde {\cal D} < {\cal D}$: 
the optimal prices of the low-$\eta^s$ strategies are positive
in all the range of $\{ h,j \}$ values for which it exists. 

In contrast, at the other side of the customers' unstability gap $[\eta_L,\eta_U]$, 
$\widetilde {\cal D}(\eta_U)={\cal D}(\eta_U)$ and since $\eta_+<\eta_U$, the 
boundary of the high-$\eta^s$ strategy (at $h=h_+(j)$) corresponds to a {\em 
negative} price. Notice that in the range $\eta_+ < \eta^s < \eta_U$ where the 
monopolist's relative maximum has a negative price, the customers demand is 
unstable and is not expected to exist at equilibrium. Thus an equilibrium with 
high demand only exists for $\eta>\eta_U(j)$. The line $h=h_U(j)$ is the 
null-price line (that is actually the line where $P=C$) on the high-$\eta$ manifold: it sets a lower bound to the 
values of $h$ for which the monopolist's high-$\eta$ strategy is viable. 

\subsubsection{Bifurcation point $B$}
In the plane $\{j,h\}$ the point $B$ defined by:
\begin{equation}
B \equiv \{j_B, h_B \equiv -\hat p_B \},
\label{eq:BB}
\end{equation}
is analogous to the point $B$ defined for the customers phase diagram (see 
Section \ref{sec:customers}). It belongs both to the line $h_+(j)$ and to the 
null-price line. The corresponding fraction of buyers is $\eta_B$. In the 
Appendix, section \ref{sec:vicinity_B} we study with some details the vicinity 
of the singular points $A$ and $B$ in the monopolist's phase diagram.

\subsubsection{Vicinity of the singular points $A$ and $B$}
\label{sec:vicinity_B}

The bifurcation point $A$ plays the 
same role, in the monopolist's phase diagram, as
the bifurcation $B$ in the Demand phase diagram \cite{NaPhGoVa05,GoNaPhSeM3as}. 
The singular behaviour at this apex $A$ is obtained 
in the very same way.
Developing in the vicinity of the $A$, at which 
$\widetilde{\Gamma}''(\eta_A)=0$, we obtain expressions 
for the boundaries of the multiple extrema region 
that are similar to those of the demand phase diagram, 
but with $\widetilde{\Gamma}$ in the place of 
$\Gamma$, $\hat \jmath$ instead of $j$ and 
$\hat \epsilon \equiv 2 \epsilon$ 
instead of $\epsilon$. 

It is interesting to consider more in details 
the vicinity of the (monopolist's) point $B$, 
where one has both $p=0$ and marginal stability 
for the `+' solution. Near $B$, for $j>j_B$ 
and/or $h>h_B$, the `+' solution gives a small 
price value, and a value of $\eta$ close to $\eta_B$. 
Like for the demand \cite{GoNaPhSeM3as}, we can expect a 
similar behaviour for the monopolist's solution: 
a linear increase of the price and a singular, 
square root, behaviour for $\eta$. Indeed at 
first non trivial order in $\epsilon$
one gets
\begin{equation}
\label{eq:nearB_+h}
\mbox{for }\; j=j_B, \; 0 < \epsilon \equiv h-h_B << 1  : \left\{
\begin{array}{ll}
p_{+} & = \epsilon, \\
\eta_{+} & = \eta_B + \sqrt{\frac{2}{\eta_B \; \Gamma'''(\eta_B)}} \; \epsilon^{1/2}, \\
\Pi_{+} & = \eta_B \; \epsilon + \sqrt{\frac{2}{\eta_B \; \Gamma'''(\eta_B)}} \; \epsilon^{3/2} \\
\end{array}
\right.
\end{equation}
And similarly,
\begin{equation}
\label{eq:nearB_+j}
\mbox{for }\; h=h_B, \; 0 < \epsilon \equiv j-j_B << 1 : \left\{
\begin{array}{ll}
p_{+} & = \eta_B \; \epsilon \\ 
\eta_{+} & = \eta_B + \sqrt{\frac{2}{\Gamma'''(\eta_B)}} \; \epsilon^{1/2} , \\
\Pi_{+} & = \eta_B^2 \; \epsilon + \eta_B \; \sqrt{\frac{2}{\Gamma'''(\eta_B)}} \; \epsilon^{3/2}  \\
\end{array}
\right.
\end{equation}
The `+' solution appears at $B$ through a 
continuous transition for the profit $\Pi$, 
with a discontinuous jump for $\eta$ (from 
$0$ to $\eta_B$), and then a square-root behaviour.
The latter is specific to the point $B$. Indeed, 
one can perform a  similar expansion in the 
vicinity of the null price line. 

Consider a point on this line with $j>j_B$. 
The corresponding value of $\eta$ is the 
solution $\eta_0(j)$ of $j=j_0(\eta)$, and 
the value of $h$ is $h_0(j)\equiv 
h_0(\eta_0(j))$. Then for 
$h=h_0(j) + \epsilon$, $0<\epsilon <<1$, expansion 
of $p=p^s(\eta)=p^d(\eta)$ at first non trivial 
order in $\epsilon$ gives
\begin{equation}
\label{eq:nearp=0}
\mbox{for }\; j > j_B, \; 0 < \epsilon \equiv h-h_0(j) << 1 : \left\{
\begin{array}{ll}
p_{+} & = \epsilon, \\ 
\eta_{+} & = \eta_0(j) + \frac{\epsilon}{\eta_0(j)\;\Gamma''(\eta_0(j))}, \\
\Pi_{+} & = \eta_0(j) \epsilon + \frac{\epsilon^2}{\eta_0(j)\;\Gamma''(\eta_0(j))}, \\
\end{array}
\right.
\end{equation}
On sees on the above expansion for $\eta$ how 
the singular behaviour at $j=j_B$ appears: 
as $j$ approaches $j_B^+$, $\eta_0(j)\rightarrow \eta_B$, 
hence $\Gamma''(\eta_0(j))$ tends to zero, and 
thus the coefficient of $\epsilon$
in the expansion of $\eta$ diverges.

\subsubsection{Correspondence between the demand and the supply branches}
\label{sec:app_lowlowhighhigh}
The fact $\eta^s_-$ and $\eta^s_+$, the low-$\eta$ and large $\eta$ solutions for the seller, do correspond to $\eta$ values falling on, respectively, the low-$\eta$ and large $\eta$ branches of the demand, is shown more formally here. 

Any solution $\eta^s_-$ corresponding to the low-$\eta$ branch for the supply lies also on the low-$\eta$ branch of the demand, because $\eta^s_- < \eta_- < \eta_L$ (see the left-hand side figures \ref{fig:examplesj35_h-12_h-14_h-15}). 

Similarly, since the zero-price line corresponds to $h_0(j)=-\hat p_U(j)$, any viable (that is with $p^s \ge 0$) solution $\eta^s_+$ on the high-$\eta$ branch for the seller also lies on the high-$\eta$ branch of the demand,  
i.e. $\eta^s_+ \geq \eta_U$.

\subsection{Monopolist's low-$\eta$ branch}
\label{sec:app_loweta}

\begin{figure}
\centering
\includegraphics[width=0.60\textwidth]{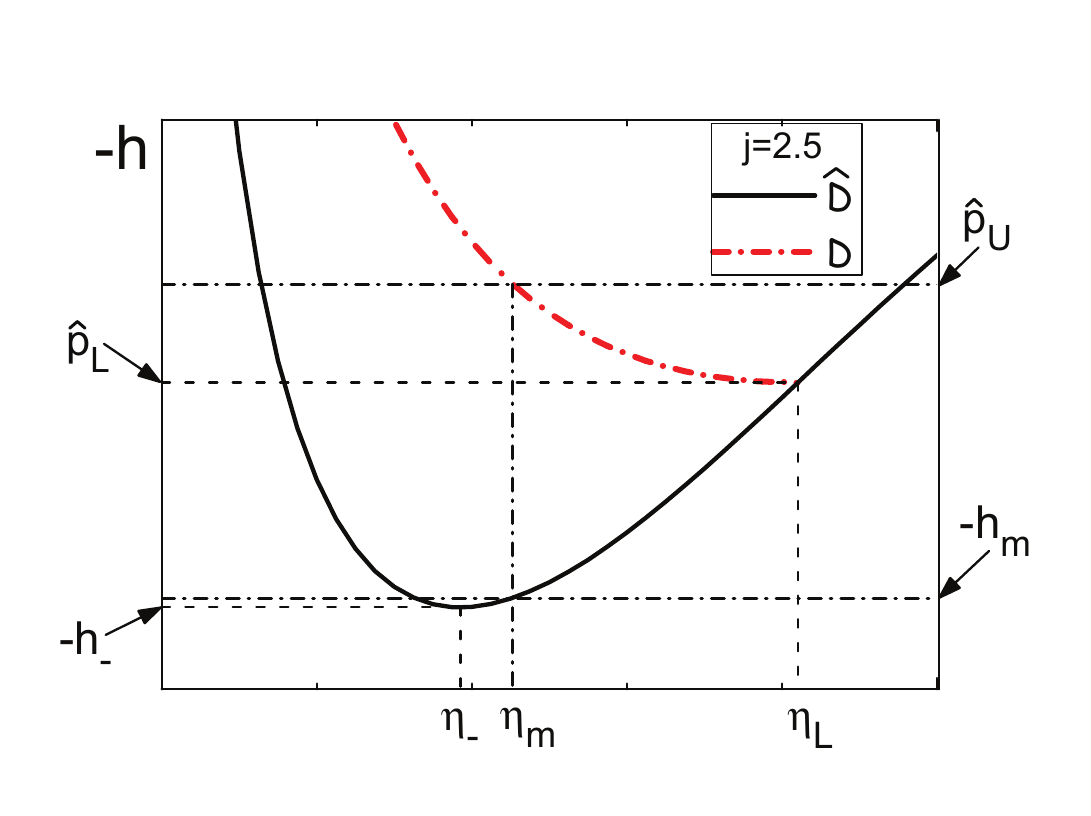}
\caption[]{\em Supply phase diagram: details of the low-$\eta$ strategy}
\label{fig:SupplyLowEta}
\end{figure}

One can make the analysis of the low-$\eta$ branch more precise. 
Let us define $\eta_m(j) < \eta_L(j)$ such that
\begin{equation}
\label{eq:eta_m}
{\cal D}(j;\eta_m)=\hat p_U(j),
\end{equation}
which is the low-$\eta$ demand at the border $\hat p_U(j)$ of the customers' multiple equilibria region. For 
$\eta_m <\eta< \eta_L$ the customers low-$\eta$ equilibrium coexists with the one at large-$\eta$. Be
\begin{equation}
\label{eq:h_m}
h_m(j) \equiv -\widetilde{\cal D}(j,\eta_m)=-\hat p_U(j)-\eta_m {\cal D}'(j;\eta_m).
\end{equation}
Depending on $j$, $\eta_m(j)$ may be smaller or larger than $\eta_-(j)$. Be $j_C (>j_B)$ the value of $j$ for which these two values of $\eta$ are equal:  $\eta_m(j_C)=\eta_-(j_C)$ and consequently $h_C \equiv h_m(j_C)=h_-(j_C)$. It verifies 
\begin{equation}
{\cal D}(j;\eta_m(j_C)) - \widetilde{\cal D}(j;\eta_m(j_C)) = \hat p_U(j_C)+h_-(j_C).
\label{eq:j_c2}
\end{equation}
In words, $j_C$ is the value of $j$ at which the (low-$\eta^s$) optimal strategy corresponds to a fraction of buyers $\eta_-$ and the corresponding price is exactly equal to $\hat p_U(j_C)+h_-(j_C) $. 

If $j> j_C$, $\eta_m < \eta_-$.  Then, for $h_m < h < h_-$ the (low-$\eta^s$) relative maximum lies at $\eta_-^s$ with $\eta_m < \eta_-^s < \eta_-$, i.e.  
inside the region with $\hat p_L < {\cal D} < \hat p_U$, where the customers' system has multiple equilibria. As shown in Appendix \ref{sec:max_boundaries}, this maximum never gives the absolute maximum of the profit, which will be on the high-$\eta$ branch. In particular one can thus conclude that the high-$\eta$ strategy becomes optimal before $h$ reaches $h_m$, that is
\begin{equation}
h_{ch}(j) < h_m(j).
\end{equation}

In addition to the $\eta_-^s$ solutions, if $h_m < - \hat p_L$ 
there exists the relative maximum at the margin $\eta_L$ of the multiple equilibria region. Although it does not correspond to an optimal pricing strategy, for completeness let us point out that 
this end point becomes a minimum for $j>j_D$ defined by
\begin{equation}
h_D \equiv h_m(j_D) = -\hat p_L(j_D),
\end{equation}
because for larger values of $j$, $h_m < - \hat p_L$.

To summarize, for $j<j_C$ and $h<h_(j)$ the low-$\eta^s$ optimal strategy
the low-$\eta^s$ strategy 
$j_C<j<j_D)$ and $h>h_m$ the profit has a 
minimum at the boundary $\eta_L$ whereas for $j>j_D$.

Figure \ref{fig:SupplyLowEta} summarizes the results for the 
low-$\eta$ manifold in the case of a logistic distribution.

\subsection{Behavior of the profit near the boundaries}
\label{sec:max_boundaries}
In addition to maxima obtained from the solutions of the 1st and 2nd order equations, as discussed above,
there may exist maxima of the profit at extreme (boundary) values: the profit
$\pi(p)$ may have a (possibly local) maximum at some of the boundary values, $p=0$, $p=\infty$,
and, in the domain of $h$ and $j$ values for which the demand $\eta^d(p)$ has two branches, 
at the maximal value $p_{U}= h +  \hat p_U(j)$ for which the high-$\eta$ demand exists,
and the minimal value $p_{L}= \max(0,\; h + \hat p_L(j) )$ for which the
low-$\eta$ demand exists. Let us consider the behaviour of the profit $\pi(p)=p \eta^d(p)$
as a function 
of the price $p$.

\subsubsection{Behaviour of the profit near $p=0$ and near $p=\infty$}
One can check that, for IWP distributions with no fat tail\footnote{In the case of a  distribution with a fat tail, one can show that for a population of $N$ customers, with $N$ large but finite, a maximum of the profit is obtained by selling a single unit of the good ($\eta=1/N$) at the customer with the largest
willingness to pay, the price being then (at least) of order $N$.} (that is with 
finite variance $\sigma$), 
the profit is decreasing to $0$
as $p$ goes to $\infty$: the large $p$ limit is always a minimum of the profit.

From the definition of the profit as $\pi(p)= p \eta^d(p)$, one has
\begin{equation}
\frac{d \;\pi}{dp} =  \eta^d(p)+ p \frac{d \;\eta^d}{dp}
\label{eq:dpi_dp}
\end{equation}
Hence at $p=0$ one has $\frac{d \;\pi}{dp} =  \eta \; > 0$: this boundary is always (unsurprisingly)
a minimum of the profit.

\subsubsection{End points of the high and low $\eta$ branches}
For $j>j_B$, the demand has two branches, $\eta^d_L$ and $\eta^d_U$, hence
the function $\pi(p)$ has itself two branches
(see Figure \ref{fig:pi_vs_p}) 
which we denote by $\pi_L$ and $\pi_U$ respectively. 
Necessarily, at least one maximum of the profit exists on each branch.

If $h + \hat p_L(j) < 0$, that is when
the low-$\eta$ branch of the demand exists already at $p= 0$, the above argument for the boundary $p=0$ applies:  
the profit  
increases along this branch when $p$ is increased from $p=0$ (going through a maximum at some $p>0$ before decreasing to zero as $p\rightarrow \infty$).

Now we know that the derivative of the demand
at each one of the boundaries,  $\hat p_{U}(j) \equiv {\cal D}(j; \eta_{U}(j))$ and $\hat p_{L}(j) \equiv {\cal D}(j; \eta_{L}(j))$, 
is singular,
$\frac{ d\eta^d(p) }{ dp }=- \infty$ (since $\frac{ dp^d(\eta) }{ d\eta }=0$, 
and the demand decreases with $p$).
It follows that, for $h > -\hat p_U(j)$, at $p_{U}= h + \hat p_U(j)$, $ \frac{d \;\pi_U}{dp} = -\infty$,
and for $h > - \hat p_L(j)$, at $p_{L}= h + \hat p_L(j)$, 
$ \frac{d \;\pi_L}{dp} = -\infty$.
Hence the boundary on the high-$\eta$ branch, $p_{U}$, is always a minimum of the profit, 
whereas for $h > - \hat p_L(j)$, $p_{L}$ is always a maximum of the profit
on the low-$\eta$ branch of the demand.  
Remark: for $h<h_m$, on the low $\eta$ branch there is a maximum for some price greater than  $p_{L}$: increasing the price from  $p_{L}$,
the profit decreases, goes through a minimum, increases up to a maximum and decreases again, going to zero as $p\rightarrow \infty$.

\subsubsection{Non optimality of the low-$\eta$ profit for $p_L\leq p \leq p_U$}
\label{sec:app_nonoptpL}
The maximum of $\pi_L$  at $p_{L}$ for $h > - \hat p_L(j)$, however, is always a (strictly) {\it local} maximum, as it is the
case for {\em any} maximum of the profit that
would exist on the low-$\eta$ branch of the demand for a price $p_L \leq p \leq p_U$.

Indeed, consider a 
value $p$ in this range, $p_L \leq p \leq p_U$. For this price the demand has two possible values, 
$\eta^d_L(d)$ on the low-$\eta$ branch (with $d = p-h$), and $\eta^d_U(d)$ on the high-$\eta$ branch. 
Since $ \eta^d_L(d) < \eta^d_U(d) $, the profit $p \eta^d_L(d)$
is strictly smaller than  $p \eta^d_U(d)$, that is the profit for the same price
obtained with the high-demand, and the later is itself smaller (or equal)
to the maximal profit associated to the high-$\eta$ branch.

In the range $h_+<h<h_-$,
an alternative way to see the fact that the profit at $p_L$ is never the
optimal one is the following, for smooth enough pdfs. 
The profit $\Pi(\eta)$ 
is a continuous function of $\eta$ with two relative maxima,
one at $\eta < \eta_-$, the other at $\eta > \eta_+$,
and a minimum in between. 
One can show that $\eta_- < \eta_L < \eta_+$, that is ${\cal D}(j; \eta)$
hits $\widetilde{\cal D}(j; \eta)$
at a value $\eta_L$ where $\widetilde{\cal D}$ is increasing 
with $\eta$. Indeed, at $\eta=\eta_L$, ${\cal D}'=0$, hence $\widetilde{\cal D}' = \eta_L {\cal D}''$,
and ${\cal D}''(j;\eta_L) = - \Gamma''(\eta_L) > 0$
($\Gamma$ is concave on $[0,\eta_B]$, see \cite{GoNaPhSeM3as}).
Since
$\eta_- < \eta_L < \eta_+$, $\eta_L$ is in between
the two maxima, so that the profit is always lower
at $\eta_L$ than at the global maximum (the largest
of the two relative maxima).

In addition, the maximum of the profit at $p_{L}$ is not a stable  
equilibrium: any small deviation of price
below the value $p_L$ would make the demand jump on the high-demand branch.

In conclusion, there is no optimal strategy for the seller corresponding to prices at the boundaries.
We have also seen that the low-$\eta$ profits, when obtained with a price in the
coexistence domain, never give an absolute maximum of the profit (see Figure \ref{fig:pi_vs_p} for an example).

\subsubsection{Non optimality of the low-$\eta$ profit at $\eta=\eta_L$}
\label{sec:app_nonoptetaL}
If  $-\hat p_L < h < h_- $,
since $\eta_- < \eta_L$, there are two intersections of $y=-h$ with $y=\widetilde{\cal D}(j; \eta)$, one at $\eta_-^s < \eta_-$ where $\widetilde{\cal D}$ has negative slope, the other at a larger value of $\eta$ where $\widetilde{\cal D}' > 0$. Only the first one corresponds to a relative maximum of the profit, the other one is a minimum. If $\eta$ is increased beyond this minimum, the profit increases continuously and reaches a relative maximum at $\eta_L$. This relative maximum does not satisfy the conditions (\ref{eq:h_monop1}) and (\ref{eq:max_cond2}) because it lies at the border of the domain $\eta \in [0,\eta_L]$ where the low-$\eta$ customers' equilibrium exists. The corresponding price is $p^s=h+\hat p_L > 0$, so that the profit is $\pi(\eta_L)=\eta_L (h+\hat p_L)$. If $\pi(\eta_L) > \pi(\eta^s_-)$ the absolute maximum of the profit lies in the high-$\eta^s$ branch, because the profit is a monotonically increasing function of $\eta$ from the minimum at $\eta<\eta_L$ up to its maximum at $\eta^s_+>\
eta_+$, reached on the high-$\eta$ branch where $\widetilde{\cal D}'(j; \eta)>0$. Conversely, if $\pi(\eta_L) < \pi(\eta^s_-)$ the absolute maximum is at either $\eta^s_-$ or $\eta^s_+$ but clearly not at $\eta_L$. In other words, the extremum at the boundary $\eta_L$ is never a winning strategy and the monopolist should never post the corresponding price.  
For smaller values of $h$, $h < - \hat p_L$ the slope of $\widetilde{\cal D}(j; \eta)$ for $\eta \in [0,\eta_L]$ is negative: there is no possible maximum at the border $\eta_L$. For these values of $h$ again there is a single optimum on the low-$\eta^s$ branch, at $\eta^s_-$.

\end{document}